\documentclass[a4page,12pt]{article}

\usepackage{amsmath,amsfonts,amssymb}
\usepackage{a4wide}
\usepackage{graphicx,graphics,color}
\usepackage[T1]{fontenc}
\usepackage[latin1]{inputenc}
\newcommand{\eps}{\varepsilon}
\usepackage{amsthm}
\usepackage{upref}

\newcommand{\sss}{\setcounter{equation}{0}}
\newtheorem{theorem}{THEOREM}[section]

\newtheorem{lemma}[theorem]{LEMMA}

\newtheorem{remark}[theorem]{REMARK}

\newtheorem{prop}[theorem]{PROPOSITION}

\newtheorem{definition}[theorem]{DEFINITION}

\newcommand{\ere}{ {\mathbb R}}

\newcommand{\CE}{{\mathbb C}}

\def\beq{\begin{equation}}
\def\ene{\end{equation}}
\def \ds {\displaystyle}
\newcommand{\bull}{\hfill $\Box$}

\def\ep{\varepsilon}
\def\by{\mathbf y}
\def\bu{\mathbf u}
\def\bv{\mathbf v}
\def\hau{\hat{u}}

\def\tu{\tilde{u}}

\def\p0{\tilde{p}_0(k,\omega)}

\begin{document}
\baselineskip=23 pt
\parskip 6 pt

\title{ Analysis of Acoustic Wave Propagation in a Thin Moving Fluid \thanks{AMS 2000 classification: 35Q72; 35Q35; 35B35; 35C15.}}
 \author{Patrick Joly and Ricardo Weder \thanks{On leave of absence from Departamento de M\'etodos
 Matem\'aticos  y Num\'ericos. Instituto de Investigaciones en Matem\'aticas Aplicadas y en Sistemas.
 Universidad Nacional Aut\'onoma de M\'exico. Apartado Postal 20-726, M\'exico
 DF 01000. Fellow of the Sistema Nacional de
Investigadores.}
\\Projet POEMS, INRIA Paris-Rocquencourt\\
 Domaine de Voluceau-Rocquencourt B.P. 105, 78153 Le Chesnay Cedex France
\\patrick.joly@inria.fr,weder@servidor.unam.mx}

\date{}
\maketitle
\begin{center}
\begin{minipage}{5.75in}
\centerline{{\bf Abstract}}
\bigskip
We study the propagation of acoustic waves in a fluid that is contained in a
thin two-dimensional tube, and that it is moving with a velocity profile that
only depends on the transversal coordinate of the tube. The governing equations are
the Galbrun equations, or, equivalently, the linearized Euler equations. We analyze the approximate model that was recently derived by Bonnet-Bendhia,
Durufl\'e and Joly to describe the propagation of  the acoustic waves in the limit when the width of the tube goes to zero.
We study this model for strictly monotonic  stable velocity profiles.
We prove that  the equations of the model of Bonnet-Bendhia, Durufl\'e and Joly  are well posed, i.e., that there is a unique global solution, and that
the solution depends continuously on the initial data. Moreover, we
prove that for smooth profiles the solution grows at most as $t^3$ as $t \rightarrow \infty$, and that for piecewise linear profiles it grows at most as
$t^4$. This establishes the stability of the model in a weak sense. These results
are obtained constructing a quasi-explicit representation of the solution. Our quasi-explicit representation gives a physical interpretation of the
 propagation of acoustic waves in the fluid and it provides an efficient way to compute numerically the solution.

\end{minipage}
\end{center}
\newpage

\section{Introduction}\sss
 \label{int}
In this paper we study the following initial value problem,

\beq \label{1.1}
\quad \left\{ \begin{array}{l} \ds
\mbox{find } u(x,y,t) : \ere \times [-1,1] \times \ere^+ \rightarrow
\ere,\\[12pt]
\ds \big(\frac{\partial}{\partial t} + M(y) \; \frac{\partial}{\partial
x}\big)^2 \; u - \frac{1}{2} \; \frac{\partial^2}{\partial x^2} \int_{-1}^1
u \; dy = 0, \quad (x,y) \in \ere
\times [-1,1], \quad t > 0,\\[18pt]
\ds u(x,y,0) = u^0(x,y), \quad (x,y) \in \ere
\times [-1,1],\\[12pt]
\ds \frac{\partial u}{\partial t} (x,y,0) = u^1(x,y), \quad (x,y)
\in \ere \times [-1,1],
\end{array} \right.
\ene where  $M \in L^{\infty}([-1,1])$  is a real-valued function,
and the initial data $u^0 \in L^2_y(H^{m+1}_x) $ and $u^1 \in
L^2_y(H^{m}_x), m=0,1, \cdots$ are given, with
\beq \label{spaces}
L^2_y(H^{m}_x) : = L^2\big([-1,1]; H^{m}(\ere) \big) \, .
\ene
We
look for solutions in ``natural energy spaces'' of the form \beq
\label{1.2} u \in C^0\big(\ere^+;  L^2_y(H^{m}_x) \big)  \cap
C^1\big(\ere^+;  L^2_y(H^{m-1}_x) \big), \quad m=0, 1, \cdots. \ene
This mathematical model has been obtained by Bonnet-Bendhia, Durufle
and Joly \cite{bdj} as an approximation in the study of a problem of
aero-acoustics. They considered the propagation of acoustic waves in
two dimensions  in a fluid that is contained in a  thin tube. The
function $M(y)$ describes, in normalized coordinates, the lateral
variations of the velocity of the fluid. Actually,  $x$ is the
coordinate along the axis of the tube and  $y$ is the transversal
coordinate. The velocity of the fluid is directed along  $x$ and it
is given by $M(y)$ at the point $(x,y)$ of the tube. The model
(\ref{1.1}) was obtained in \cite{bdj} by a formal asymptotic
expansion  on the width  of the tube of the solution to Galbrun
equations \cite{gal}, that are equivalent to the linearized Euler
equations. Physically, the validity of our model requires that the
transverse dimension of the tube is small with respect to the
wavelength but no too small to justify the fact of neglecting the
viscosity effects : in this sense, this model can be seen as a low
frequency model. Another potential application of this study is the
construction of effective boundary conditions - also called lining
models - to take into account the modeling of boundary layers in
aeroacoustics, which is a quite delicate issue from mathematical and
numerical points of view (see for instance \cite{Brambley}).

In \cite{bdj} the analysis of equation (\ref{1.1}) was reduced to a one
dimensional problem  by Fourier transform along $x$, taking advantage of
the fact that the velocity profile  is only a function of the transversal
coordinate $y$. This reduces the study of the solutions to (\ref{1.1}) to the
spectral theory of a non-local operator, $A$, that acts only on the transversal
variable $ y \in [-1,1]$:  $A \in {\cal L}\big(L^2[-1,1])^2\big)$. More precisely, setting
$$
U(x,y,t) := \big( \, u(x,y,t), (\partial_t + M(y) \partial_x)u(x,y,t) \, \big)^t
$$
and denoting by $ \hat{U}(k,y,t)$ the Fourier transform along $x$ of
$U(x,y,t)$ one has formally
\begin{equation} \label{expA}
\frac{d\hat{U}}{dt}(k, \cdot ,t) + i \, k \, A \, \hat{U}(k, \cdot
,t) = 0 \quad \Longrightarrow \quad \hat{U}(k, \cdot ,t) = e^{-i \,
k \, A \, t} \; \hat{U}(k, \cdot ,0).
\end{equation}
It was shown in \cite{bdj} that on spite of its
apparent simplicity this problem has rather surprising properties. As the
operator $A$ is not normal, it can have complex (non-real) eigenvalues. It was pointed out
in \cite{bdj} that a necessary
condition in order that  the problem  (\ref{1.1}) is well posed and stable - in the sense
that the solutions do not grow exponentially in time- is that all the
eigenvalues of $A$ are real. These issues were studied in detail in \cite{bdj}.
General properties of  $A$ were obtained and the general structure of the point
and continuous spectrum  were analyzed. Moreover, several results on the
existence and on the absence of complex eigenvalues of $A$ were given. These results were illustrated numerically. Furthermore, it
was conjectured in \cite{bdj}
that the condition that $A$ has no complex eigenvalues is also sufficient for
the stability of the problem (\ref{1.1}). Unfortunately, it does not seem that
such a result can easily be deduced from the standard semi-group theory.

That is why, in this paper, we take a point of view that is slightly different from the one of
\cite{bdj}.  Actually, we directly  solve  equations (\ref{1.1})   by
Fourier-Laplace  transform. As usual, to prove  that the solution does not grow
exponentially in time it is necessary to deform the integration contour of the
inverse Laplace transform to the real axis. This requires that the
norming coefficient, $N(\lambda)$, in the inverse Laplace transform (see
(\ref{3.7})) has no complex poles.
It turns out that the poles of $N(\lambda)$  in  $\mathbf C \setminus \hbox{\rm
Range}\, M$  are precisely the eigenvalues of
the operator $A$ of \cite{bdj} in  $\mathbf C \setminus \hbox{\rm
Range}\, M$, what means that we obtain, as to be expected, the same stability
condition as in \cite{bdj}. Moreover, $N(\lambda)$ has, in general, a cut on $ \hbox{\rm
Range}\, M$ and, furthermore, it can blow up as we approach the cut from above and from below, what makes
the issue of deforming the contour to the real axis on both sides of the cut
quite delicate. Although we think that our approach is quite general, it is
difficult to state general results.  The method that we present here to construct a quasi-explicit representation of the solution and to prove  stability
can be applied to general profiles, provided that one can analyze the limiting values of the norming factor $N(\lambda)$ as $\lambda$ tends to
$[M_-,M_+]$ from above and from below. In this paper, we develop the
well-posedness theory in two
cases. First, we consider strictly monotonic and convex (or concave), smooth
profiles. In this case, we prove that the norming factors $N(\lambda)$ of our profiles have continuous limiting values as
 we approach the cut from above and from
below, that are different. We use this result to prove that the
problem (\ref{1.1}) has   a unique solution in the natural class of
fields (\ref{1.2}). Moreover, we prove that the solutions grow at
most as $t^3$ as $t \rightarrow \infty$. Then, we consider the case of
piecewise linear profiles for which the structure of $N(\lambda)$ is
quite different. In our mind, the interest of this class of profiles
(already considered in \cite{bdj} as a theoretical tool) is to
provide a safe way to get numerical approximations of the solution. We also  prove that for these profiles the
problem (\ref{1.1}) has   a unique solution in the natural class of
fields (\ref{1.2}). However, in this case we prove that the solutions grow at
most as $t^4$ as $t \rightarrow \infty$. These results establish that the
model is stable in a weak sense,  and   prove the conjecture of
\cite{bdj} for our class of profiles.
For both cases of profiles our method, that is
based in the Fourier-Laplace transform, leads to a quasi-explicit
representation of the solution, that gives a physical interpretation
of the propagation of acoustic waves in the fluid, and it provides an
efficient way to compute numerically the solution. The numerical
results are presented in \cite{jjw}.

The paper is organized  as follow. In Section 2 we briefly state,
for the reader's convenience, a derivation of the approximate model
(\ref{1.1}) from the Galbrun equations \cite{gal} that is slightly
different from the one  in \cite{bdj}. In Section 3 we develop the
well-posedness theory and we prove our results on the existence and
uniqueness of solutions in the space of
 fields (\ref{1.2}) and on the continuous dependence on the
initial data for the two classes of  profiles (Subsections
\ref{sec3.3.1} and \ref{sec3.3.2}, respectively). Finally, in
Section 4, we obtain a quasi-explicit representation of the
solutions to (\ref{1.1}), of which we give a physical
interpretation.

\section{ Derivation of the quasi-1D model}\sss
We model this flow by means of the equations of Galbrun \cite{gal}
where the unknown $(\bu_{\ep},\bv_{\ep})$, that are functions of $(x,\by,t)$,
are the components of the Lagrange displacement of the
fluid. The governing
equations are,
\beq \label{2.3}
\left\{ \begin{array}{ll}
\ds \Big(\frac{\partial }{\partial t} + M_{\ep}(\by) \;
\frac{\partial}{\partial x}\Big)^2  \bu_{\ep} - \frac{\partial}{\partial
x}
\left( \frac{\partial \bu_{\ep} }{\partial x} + \frac{\partial
    \bv_{\ep}}{\partial
    \by} \right) = 0, & (x, \by) \in \Omega_{\ep}, \quad t > 0, \\[18pt]
\ds  \Big(\frac{\partial }{\partial t} + M_{\ep}(\by) \;
\frac{\partial}{\partial x}\Big)^2  \bv_{\ep}
- \frac{\partial}{\partial \by}
\left( \frac{\partial \bu_{\ep}}{\partial x} + \frac{\partial
    \bv_{\ep}}{\partial \by}\right) = 0, & (x, \by) \in \Omega_{\ep}, \quad t > 0,
\end{array} \right.
\end{equation}
with the boundary condition that on the walls of the tube,  $\by = \pm \;
\ep$, the normal component of the displacement is zero,
\beq \label{2.4}
\bv_{\ep}(x , \pm \; \ep, t) = 0, \quad x \in \ere, \quad t > 0.
\ene
To determine formally the approximate model (see  \cite{bdj}) let us first perform a change
of scale in  $\by$ ($ y:= \by /\, \ep   $) to work in a fixed domain and use a
scaling of the unknown $\bv_{\eps}$ (which we expect to tend to 0 since this
function is defined on a thin domain and is imposed to $0$ on the boundary of
this thin domain).
Let us denote,
\begin{equation} \label{2.5}
\left\{ \begin{array}{ll}
\ds u_{\ep}(x,y ,t):=\bu_{\ep}(x, \ep y,t), & (x, y):= \in \ere \times (-1,1)
, \quad t > 0, \\[12pt]
\ds   v_ {\eps}(x,y,t):= \bv_{\eps}(x,\ep  y,t) / \eps  ,& (x, y) \in \ere \times (-1,1)
, \quad t > 0.
\end{array} \right.
\end{equation}
The equations in  $(u_{\ep}, v_{\ep})$ are given by,
\beq \label{2.6}
\left\{ \begin{array}{ll}
\ds \Big(\frac{\partial }{\partial t} + M(y) \;
\frac{\partial}{\partial x}\Big)^2  u_{\ep} - \frac{\partial}{\partial
x}
\left( \frac{\partial u_{\ep} }{\partial x} +
  \frac{\partial v_{\ep}}{\partial
    y} \right) = 0, & (x, y) \in \ere \times (-1,1), \quad t > 0, \\[18pt]
\ds  \, \eps² \, \Big(\frac{\partial }{\partial t} + M(y) \;
\frac{\partial}{\partial x}\Big)^2  v_{\ep}
- \frac{\partial}{\partial y}
\left( \frac{\partial u_{\ep}}{\partial x} + \frac{\partial
    v_{\ep}}{\partial y}\right) = 0, & (x, y) \in \ere \times (-1,1), \quad t > 0,
\end{array} \right.
\ene
with the boundary conditions,
\beq \label{2.7}
v_{\ep}(x , \pm \; 1, t) = 0, \quad  x \in \ere, \quad t >0.
\ene
We obtain the limit model by dropping the $\eps^2$ term in the second equation
of (\ref{2.6}) : the formal limit $(u,v)$ of $(u_{\ep},v_{\ep})$ satisfies
\beq \label{2.6bis}
\left\{ \begin{array}{ll}
\ds \Big(\frac{\partial }{\partial t} + M(y) \;
\frac{\partial}{\partial x}\Big)^2  u - \frac{\partial}{\partial
x}
\left( \frac{\partial u}{\partial x} +
  \frac{\partial v}{\partial
    y} \right) = 0, & (x, y) \in \ere \times (-1,1), \quad t > 0, \\[18pt]
\ds \frac{\partial}{\partial y}
\left( \frac{\partial u}{\partial x} + \frac{\partial
    v}{\partial y}\right) = 0, & (x, y) \in \ere \times (-1,1), \quad t > 0.
\end{array} \right.
\ene
Then, there is a function of $x,t$ only, that we denote by  $p(x,t)$, such that,
$$
\Big( \frac{\partial u}{\partial x} + \frac{\partial v}{\partial y}
\Big)(x,y,t) = -p(x,t).
$$
Integrating in $y$ and as  $v(x , \pm \; 1, t) = 0$,
we prove that,
\beq
\label{2.9}
p(x,t) = -\frac{1}{2} \; \int_{-1}^1 \frac{\partial u}{\partial x}(x, y, t)
\; dy.
\ene
The first equation (\ref{2.6}) writes
$$
\Big(\frac{\partial }{\partial t} + M(y) \;
\frac{\partial}{\partial x}\Big)^2  u - \frac{\partial}{\partial x} \left(
\frac{\partial u}{\partial x} + \frac{\partial v}{\partial y} \right)
\equiv \Big(\frac{\partial }{\partial t} + M(y) \;
\frac{\partial}{\partial x}\Big)^2  u + \frac{\partial p}{\partial x}= 0.
$$
Finally, replacing $p$ by its value given by (\ref{2.9}) we obtain
equation (\ref{1.1}).

\section{ Well-posedness of the quasi-1D model}\sss
\label{section3}
\subsection{Preliminary material}
In what follows we assume that $M(y)$ is either increasing or
decreasing and we denote,
$$
M_+:=  \sup_{\ds y \in (-1,1)} M(y),\quad  M_-:= \inf_{\ds y \in (-1,1)}  M(y).
$$
 We write (\ref{1.1}) as follows,
\beq
\big(\frac{\partial}{\partial t} + M(y) \; \frac{\partial}{\partial
x}\big)^2 \; u(x,y,t) =  \frac{\partial^2}{\partial x^2}  \big[{\bf a}(u)\big](x,t)
\label{3.1},
\ene
where $v\rightarrow {\bf a}(v)$ is the averaging operator in the $y-$direction:
\beq
\big[{\bf a}(v)\big](\cdot):=  \frac{1}{2} \; \int_{-1}^1 v(\cdot,y)\; d y.
\label{3.2}
\ene
From (\ref{3.2}), it is clear that, if ${\bf a}(u)$ is known, (\ref{3.1})  is a
simple transport square equation along the axis of the tube for each fixed $y$, with a $y$-dependent
transport velocity   that is given by the profile $M(y)$. We solve this
transport equation explicitly. In this way  we
obtain a quasi-explicit representation of the unique solution  to
(\ref{1.1}). Therefore, we would like to get an equation for the average value
 ${\bf a}(u)$ alone. As we shall see, such an equation in  ${\bf a}(u)$ is obtained by applying the Fourier
 transform in $x$ and the Laplace transform in $t$. Then, the well-posedness of (\ref{1.1})
 is reduced to showing that it is possible to apply the inverse Fourier-Laplace
 transform,  and  also to obtain a priori estimates for ${\bf a}(u)$.\\[12pt]
As usual, we define the Fourier transform as an unitary operator on
$L^2(\ere)$, defined for $f \in L^1(\ere)$ as
$$
(\mathcal F f)(k)=\hat{f}(k):= \frac{1}{\sqrt{2\pi}}\, \int_{-\infty}^\infty\, e^{-ikx}\, f(x)\,
dx.
$$
After Fourier transform equation (\ref{3.1}) becomes (this is nothing but
(\ref{expA}) written is scalar second order form),
\beq
\label{3.3}
\ds \big(\frac{\partial}{\partial t} + M(y) \;ik \big)^2 \; \hau = -k^2  {\bf a}(\hau)(k,t), \quad k \in \ere, \quad t > 0.
\ene
We take a point of view that is slightly different from the one of
\cite{bdj}.  Instead of writing  equation  (\ref{3.3})  as an evolution problem for a first order system, and
 using semigroup theory to reduce the problem  to the spectral analysis of the
 generator $A(k) \equiv k \, A$ of the system,
as it was  done in \cite{bdj}, we directly  solve  equation (\ref{3.3}) by
Laplace transform in time.\\[12pt]
We denote by $\tu$ the Laplace transform in time of $\hau$,
\beq \label{3.4}
\tu (k,y,\omega ):= \int_0^\infty\, e^{i\omega t} \;  \hau (k,y,t)\, dt, \quad
\Im \omega > 0.
\ene
\begin{remark} One easily checks that, for each $k \in \ere$,
  (\ref{3.3}) admits a unique solution $\hau(k, \cdot, t)$ which increases in time at most exponentially as (because $A$ is bounded
  \cite{bdj}, see (\ref{expA}))
$$
\exp{B|k|t} \quad \mbox{with } B := \| A \|> 0,
$$
so that $\hau(k, \cdot, t)$ has a well defined Laplace transform as
soon as $\Im \, \omega > B |k|$.
\end{remark}
\noindent After Laplace transform, (\ref{3.3}) becomes,
\beq \label{3.5}
\ds- \big(\omega - M(y) \; k \big)^2 \; \tu +k^2  {\bf a}(\tu)(k,t) = \hau^1- i
(\omega -2 k M(y))\hau^0,
\ene
where $u(x,y,0)= u^0(x,y)$, and $ \ds \frac{\partial}{\partial t} u(x,y,0)=
u^1(x,y)$.
Dividing both sides of (\ref{3.5}) by $(\omega - k M(y))^2$ and taking the average
over $y$ of both sides    we prove that,
\beq
\label{3.6}
{\bf a}(\tu )= -2 N\left(\frac{\omega}{k}\right)\, {\bf a}\left( \big[\hau^1 -i (\omega -2k M(y))
    \hau^0\big ]\, \big[\omega -k M(y)\big]^{-2} \right),
\ene
where $N$ is the norming factor,
\beq \label{3.7}
N(\lambda):=  \big(2-
 F(\lambda)\big)^{-1},
\ene
 with,
\beq
\label{3.8}
 F(\lambda):=
\int_{-1}^1\, (M(y)-\lambda)^{-2} \, dy, \quad \lambda \in \CE\setminus [M_-,M_+].
\ene
Inverting the Laplace transform  and changing the integration variable $\omega$
to $\lambda= \lambda_R+i \lambda_I:= \omega/k$, we obtain the following
representation for  ${\bf a}(\hau)$,
\beq
\label{3.9}
{\bf a}(\hau)(k,t)= \hat{\bf a}_0(k,t) + \hat{\bf a}_1(k,t)
\ene
where $\hat{{\bf a}}_0(k,t)$ and $\hat{{\bf a}}_1(k,t)$ are,  respective, the contributions of
the Cauchy data $u^0$ and $u^1$ and are given by:
\beq
\label{3.9b}
\hat{\bf a}_j(k,t) = -\frac{k^{-j}}{\pi}\,\int_{\ds \pm \ere+i \lambda_I}\, e^{-ik \lambda
t} \, N(\lambda)\, \, {\bf a}\big(\hat{f}_j(\cdot,\lambda) \, \hau^j(k,\cdot)\big)\, d\lambda,
\ene
where $\lambda_I > 0$  for $ k >0$ and  $ \lambda _I < 0$ for $k<0$ with
$|\lambda_I|$ large enough. Furthermore,
\beq
\label{3.10}
\hat{f}_0(y,\lambda):=  i M(y)\,
(\lambda - M(y))^{-2}-  i  \, (\lambda - M(y))^{-1}, \quad
\hat{f}_1(y,\lambda):= (\lambda - M(y))^{-2}.
\ene
\\[0pt]
For proving the well-posedness of (\ref{1.1}), our objective is to derive a
priori estimates for $\hat{\bf a}_0(k,t)$ and  $\hat{\bf a}_1(k,t)$ as functions
of $k$ and $t$.
Such estimates can not be deduced directly from ( \ref{3.9b}, \ref{3.10})
as if $\lambda$ is not real, the exponent in  the
right-hand side of (\ref{3.9b}) will blow up as $k \rightarrow \pm \infty$,
unless the initial data decays very fast as $|k| \rightarrow \infty$, and  we
will not be able to invert the Fourier transform to compute $\big[{\bf
  a}(u)\big](x,t)$.
It is now imperative to transform the expressions (\ref{3.9b}, \ref{3.10}) using complex
integration techniques (contour deformation). This is the object of the next
subsection.
\subsection{A new expression for $\hat{\bf a}_0(k,t)$ and  $\hat{\bf a}_1(k,t)$}
We consider the case $ k >0$. For $ k <0$ the results are obtained in the same way, with obvious changes.

According to (\ref{3.9b}) and (\ref{3.10}), we simply need to compute the integrals
\beq \label{3.11}
I_\ell(kt,y) :=  \int_{\ere+i \lambda_I}\, {e^{-ik\lambda t}} \;
\frac{N(\lambda)}{\big(\lambda-M(y)\big)^{\ell+1}} \;\, d \lambda, \quad \lambda_I >0,
\quad \ell = 0,1.
\ene
Indeed,  we immediately obtain from ( \ref{3.9b}, \ref{3.10})
\beq \label{ha0}
\hat{\bf a}_0(k,t) = \frac{-i}{\pi} \;
{\bf a}\Big(\big[M(\cdot) I_1(kt,\cdot) - I_0(kt,\cdot)\big] \, \hau^0(k,\cdot)
\Big),
\ene
\beq \label{ha1}
\hat{\bf a}_1(k,t) = \frac{-1}{\pi k} \;
{\bf a}\Big( I_1(kt,\cdot)  \, \hau^1(k,\cdot)
\Big).
\ene
We are going to compute the integrals (\ref{3.11}) by using complex integration methods. For this, we need to describe in detail the
analyticity properties of $N(\lambda)$. From now on, we shall assume that,
$$
({\cal H}_1) \quad \mbox{$M$
is a continuous function of $y$.}
$$
We set
\beq \label{defM-M+}
M_- = \min_{y \in [-1,1]} M(y), \quad M_+ = \max_{y \in [-1,1]} M(y).
\ene
From formula (\ref{3.8}), it is clear that $F$ is analytic in $\CE \setminus [M_-, M_+]$ so that
$N$ is meromorphic in $\CE \setminus [M_-, M_+]$. Observe that as
$$
\lim_{|\lambda|\rightarrow \infty} N(\lambda)= \frac{1}{2}
$$
the poles of $N(\lambda)$ are contained in a bounded set of $\CE$. Since
$N(\overline{\lambda}) = \overline{N(\lambda)}$, we deduce that these poles are
equally distributed with respect to the real axis. At this point, we need to
make a fundamental assumption
$$
({\cal H}_S) \quad \mbox{The function $N(\lambda)$ has no complex (i.e. non
  real) poles.}
$$
This assumption is clearly related to the velocity profile $M$. The paper \cite{bdj}  gives explicit conditions on $M$ that ensure that $({\cal
  H}_S)$ is satisfied. We shall give another proof of such a result in Lemma \ref{le3.1}.
  The paper \cite{bdj} also gives examples of profiles for which $({\cal H}_S)$ does
  not hold.
We shall say
that a profile $M$ is {\bf stable} when the assumption $({\cal H}_S)$ is
satisfied. This denomination is justified by the fact that one easily proves
that if $({\cal H}_S)$ does not hold, the Cauchy problem is strongly {\bf
  ill-posed} (see \cite{bdj} - the complex poles of $N(\lambda)$ are precisely
the complex eigenvalues of the
operator $A$ of \cite{bdj}). The assumption $({\cal H}_S)$ is thus a {\bf necessary
condition} for the well-posedness of (\ref{1.1}). Our conjecture is that this condition is
{\bf also sufficient}. The object of this paper is to show that this conjecture is
true under additional technical assumptions on $M$.
\\[12pt]
Concerning the real poles of $N(\lambda)$, we note (see again  \cite{bdj}
for details) that $N(\lambda)$ has at most one pole in  $(-\infty, M_- )$ that
we
denote by $\lambda_-$, if it exists, and at most one pole in $ (M_+, \infty)$
that we denote by $\lambda_+$ if it exists, and that these poles are simple.
This is immediate since,
$$
\frac{\partial}{\partial \lambda}\, F(\lambda)= -2 \int_{-1}^1
\frac{1}{(M(y)-\lambda)^3}   d \lambda
$$
is positive for $ \lambda > M_+$ and  is negative for $ \lambda <  M_-$.
Furthermore, the residues of $N(\lambda)$ at the poles are given by,
\beq
\label{3.14}
\mbox{Res }N(\lambda_\pm)= \frac{-1}{F'(\ds\lambda_\pm)}.
\ene
We make the additional assumption,
$$
({\cal H}_{RL}) \quad \mbox{ $M(y)$ takes the value $M_\pm$ at a point where it has a right or a left
 derivative. }
$$
One easily sees (see also \cite{bdj}) that in this case
$F(\lambda)$ blows up to $+ \infty$ when
 $\lambda$ approaches $M_\pm$, from real values, which ensures the existence of
 $\lambda_\pm$.

\begin{figure}[h]
\centerline{
\includegraphics[width=10cm]{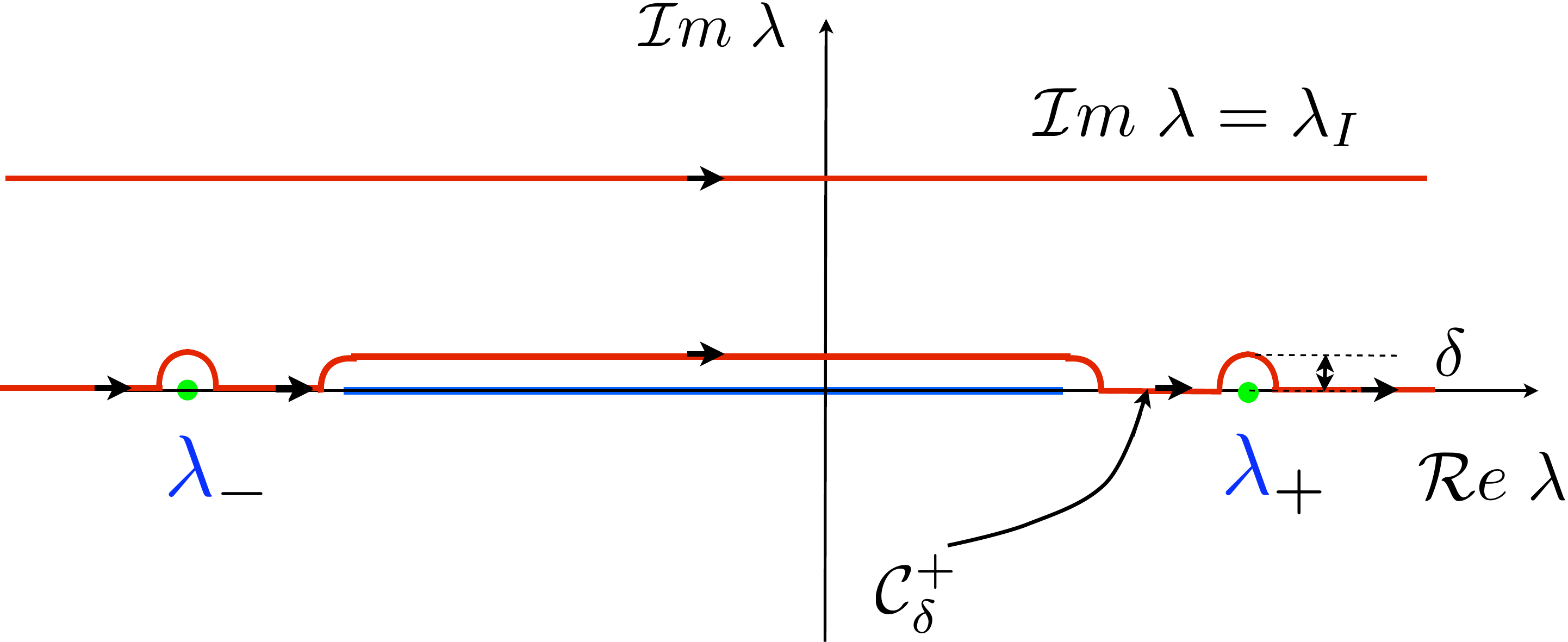}}
\caption{The integration contour ${\cal C}_{\delta}^+$}
\label{Contour1}
\end{figure}

 We now present  our computations of $I_{\ell}$ in the case $k
>0$. As mentioned above, for $k <0$ the results are obtained in the same way. First, thanks to $({\cal H}_S)$, we can deform the
line $ {\cal I}m \;\lambda = \lambda_I$ into the contour ${\cal
C}_{\delta}^{+}$ that coincides with the real axis outside a
$\delta$-neighborhood of the poles $\lambda_{\pm}$ and the cut
$[M_-,M_+]$ (see Figure \ref{Contour1}) to obtain (note that
${N(\lambda)}/{\big(\lambda-M(y)\big)^{\ell+1}}$  decays to 0
when $|{\cal R}e\; \lambda|$ goes to infinity):

\beq \label{3.12}
I_\ell(kt,y) =  \int_{{\cal C}_{\delta}^{+}}\, {e^{-ik\lambda t}} \;
\frac{N(\lambda)}{\big(\lambda-M(y)\big)^{\ell+1}}\; \, d \lambda.
\ene
\vspace{1cm}

\begin{figure}[h]
\centerline{
\includegraphics[width=10cm]{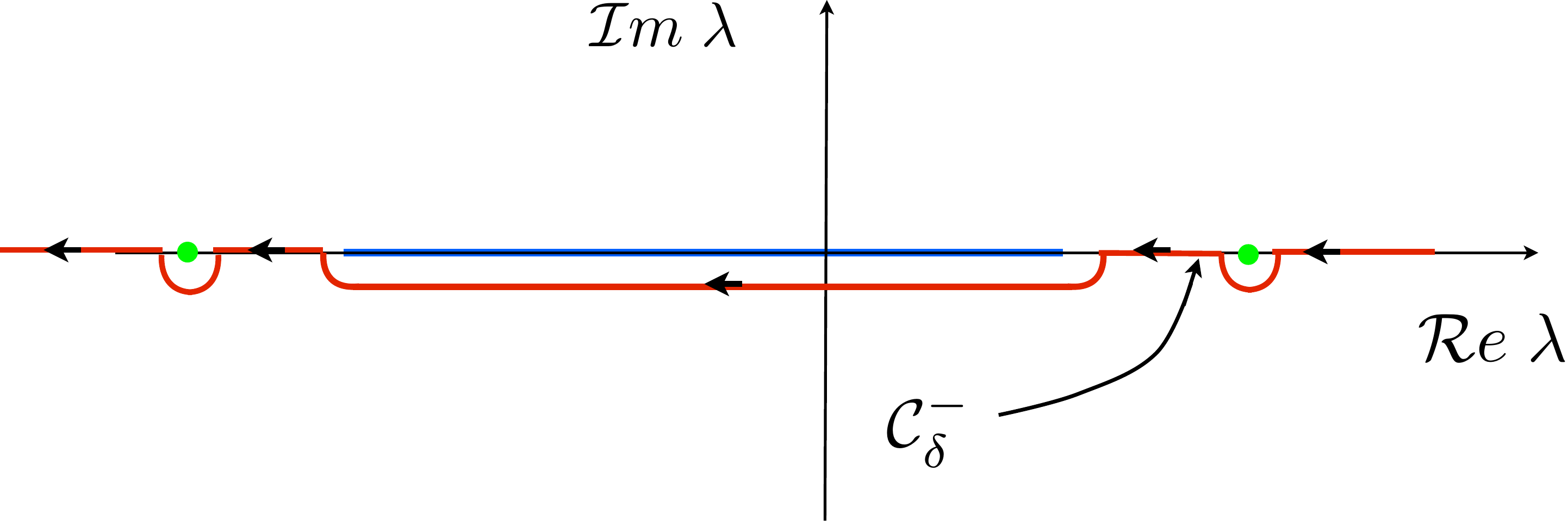}}
\caption{The integration contour ${\cal C}_{\delta^-}$}
\label{Contour2}
\end{figure}

Let ${\cal C}_{\delta}^{-}$ be the contour deduced from
${\cal C}_{\delta}^{+}$ by
symmetry with respect to the real axis (see Figure \ref{Contour2}).

\begin{figure}[h]
\centerline{
\includegraphics[width=10cm]{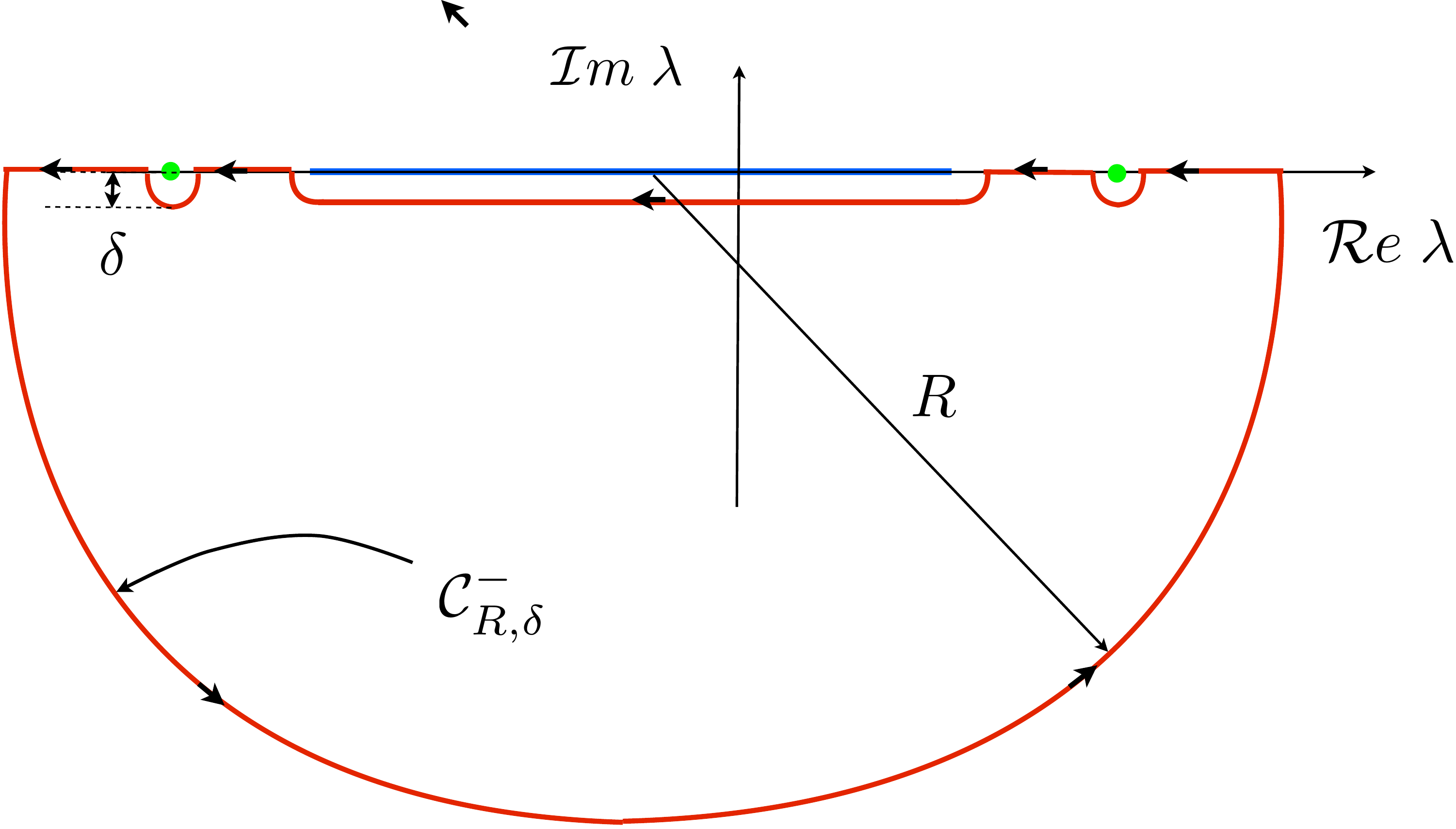}}
\caption{The integration contour ${\cal C}_{R, \delta}$}
\label{Contour3}
\end{figure}

We claim that,
\beq \label{3.13}
\int_{{\cal C}_{\delta}^{-}}\, {e^{-ik\lambda t}} \;
\frac{N(\lambda)}{\big(\lambda-M(y)\big)^{\ell+1}}\; \, d \lambda = 0.
\ene
This is obtained by considering the closed contour ${\cal C}_{R,\delta}^{-}$ of
Figure \ref{Contour3} inside which the integrand is analytic.

\begin{figure}[h]
\centerline{
\includegraphics[width=10cm]{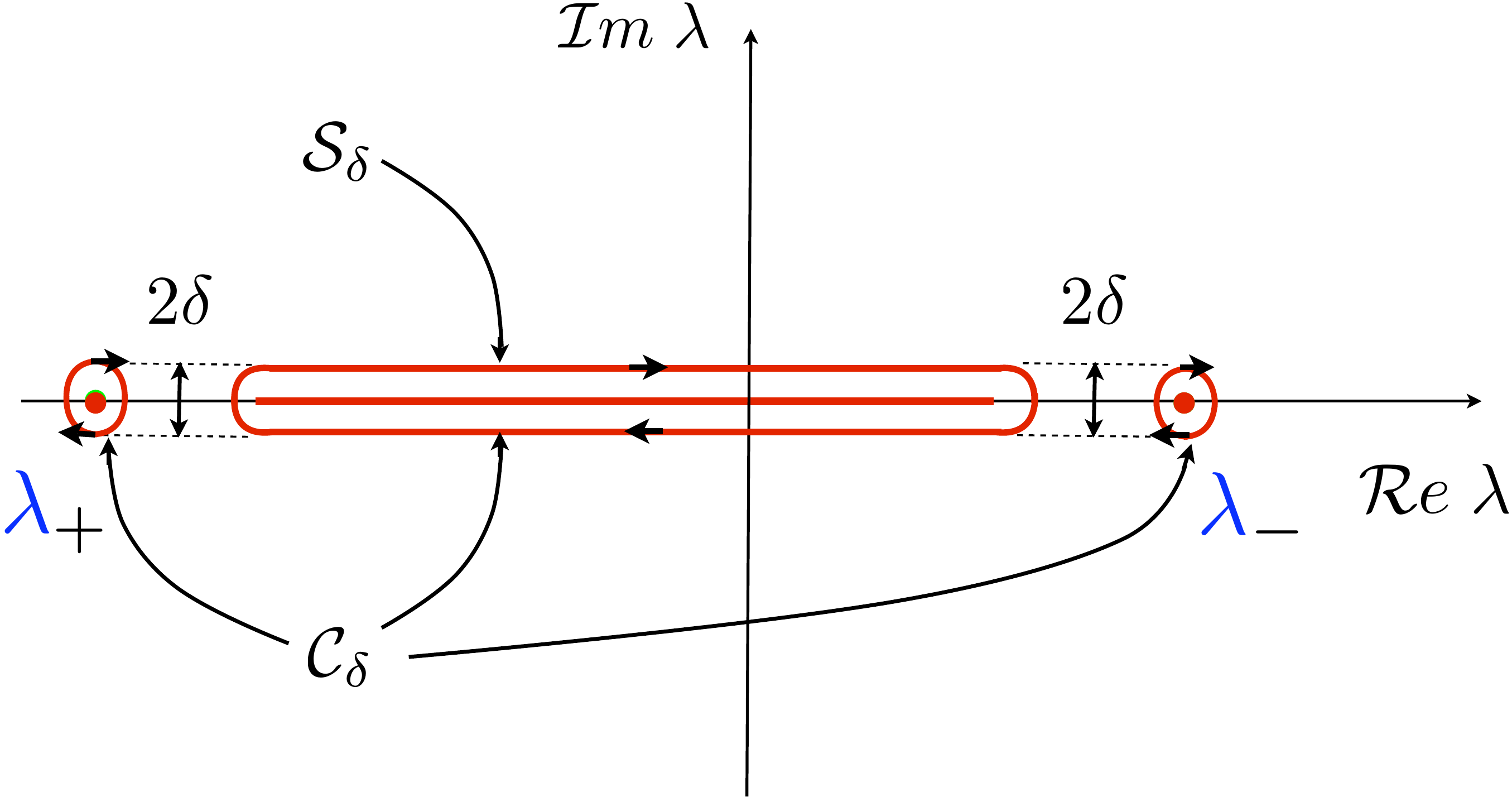}}
\caption{The integration contours ${\cal C}_{\delta}$ and ${\cal
S}_{\delta}$} \label{Contour4}
\end{figure}

\noindent  Thus, by
Cauchy's theorem
\beq \label{3.14b}
\int_{{\cal C}_{R,\delta}^{-}}\, {e^{-ik\lambda t}} \,
\frac{N(\lambda)}{\big(\lambda-M(y)\big)^{\ell+1}}\; \;d \lambda = 0.
\ene
Then, (\ref{3.13}) is obtained from (\ref{3.14b}) by passing to the limit when
$R \rightarrow + \infty$: the contribution of the integral along the
semi-circle of radius $R$ vanishes because it is inside the good complex
half-space when $k > 0$ (for the exponential term) and because the function
${N(\lambda)}/{\big(\lambda-M(y)\big)^{\ell+1}}$ decays at infinity.
\\[12pt]
Finally adding (\ref{3.12}) and (\ref{3.13}), we obtain
\beq \label{3.11b}
I_\ell(kt,y) =  \int_{{\cal C}_{\delta}}\, {e^{-ik\lambda t}} \;
\frac{N(\lambda)}{\big(\lambda-M(y)\big)^{\ell+1}}\, d \lambda,
\ene
where ${\cal C}_{\delta}$ is the closed contour of Figure
\ref{Contour4}.

\vspace{1cm}

By the theorem of residues we also have:
\beq
\label{3.19}
I_\ell(kt,y)= I_{\ell,p}(kt,y)+I_{\ell,c}(kt,y),\quad  \ell=1,2,
\ene
where
\beq
\label{3.21}
\left\{ \begin{array}{l}\ds
\displaystyle I_{\ell,p}(kt,y):= -  2 i \pi \sum_{\pm} \; e^{-ik\lambda_\pm t} \,
  \frac{ \mbox{Res}\, N(\lambda_\pm)}{(\lambda_\pm-M(y))^{\ell+1}},
\\[18pt]
\displaystyle I_{\ell,c}(kt,y) :=  \int_{{\cal S}_{\delta}}\, {e^{-ik\lambda t}}
\; \frac{N(\lambda)}{\big(\lambda-M(y)\big)^{\ell+1}} \; d \lambda,
\end{array} \right.
\ene
where ${\cal S}_{\delta}$ is a closed curve that collapses to the segment
$[M_-, M_+]$ when $\delta$ goes to 0 (see Figure
\ref{Contour4} again).

To get a more exploitable formula for
$I_{\ell,c}(kt,y)$ , we would like to take the limit of the expression in the
right-hand side when $\delta$ tends to 0. For this we need some additional
information on the behaviour of $N(\lambda)$ when $\lambda$ approaches the
segment $[M_-, M_+]$.
In the following we consider two different situations leading to quite
different behaviours of $N(\lambda)$ when $\lambda$ approaches $[M_-, M_+]$.
\begin{itemize}

\item The case of a class of smooth profiles with constant monotonicity and
  convexity (see Subsection \ref{sec3.3.1}). This is a physically relevant case
  for which $N(\lambda)$ has two distinct limits on $[M_-, M_+]$ depending on
  whether one approaches the segment from above or from below.
  In this case, the function $N(\lambda)$ presents a cut along $[M_-, M_+]$.

\item The case (already considered in \cite{bdj}) of a class of piecewise
  linear profiles (see Subsection
  \ref{sec3.3.2}), typically with fixed
  convexity . This case is interesting to
  provide a safe procedure for numerical approximations. In this situation, the
  function $N(\lambda)$ is meromorphic (it is a rational fraction) with real
  poles distributed in $[M_-, M_+]$, as well as poles at $\lambda_\pm$.
\end{itemize}
\subsection{Well-posedness results}
\subsubsection{For a class of smooth profiles} \label{sec3.3.1}
We denote by $C^{(n,\gamma)}[-1,1], n=0,1,\cdots,  0< \gamma < 1, $ the space of real-valued functions defined on $[-1,1]$ that are $n-$times
differentiable and with the derivative of order $n$ H\"older continuous with exponent $\gamma$.

\begin{definition} \label{Cs}
If $l$ denotes an integer $\geq 2$ and $\gamma$ a real number in $]0,1[$ , we
shall say that a profile $M(y)$ belongs to the class ${\cal C}_s^{l,\gamma}$
if it fulfills the following conditions:
\beq
M \in C^{(l,\gamma)}[-1,1], \quad
M'(y) \neq 0, \quad M''(y) \neq 0, \quad \forall \; y \in [-1,1].
\ene
In other words, a profile in the class ${\cal C}_s^{l,\gamma}$ is smooth, strictly
monotonous and
has a fixed (strict) convexity.
\end{definition}
\begin{lemma}
\label{le3.1}
Suppose that $M(y)$ belongs to the class ${\cal C}_s^{l,\gamma}$ for some $l
\geq 2$ and $ \gamma \in ]0,1[$. Then, it is stable in
the sense that it satisfies the condition $({\cal H}_S)$.
\end{lemma}
\noindent {\it Proof:} Without loss of generality, we can assume that $M$ is
increasing (changing $M$ into $-M$ simply exchanges the signs of the roots of
$F(\lambda) = 2$) and that $M_-\big(\equiv M(-1)\big)=0$
(translating $M$ simply translates the solutions of $F(\lambda) = 2$ parallel to
the real axis). Since, by continuity, $M'(y)$ is bounded from below by a
strictly positive constant, we can write
$$
F(\lambda) = \int_{-1}^1 \frac{M'(y)}{(M(y) - \lambda)^2} \; \frac{dy}{M'(y)}
= - \int_{-1}^1 \frac{d}{dy}\left\{\frac{1}{(M(y) - \lambda)} \right\}\; \frac{dy}{M'(y)}.
$$
After integration by parts, we get
$$
F(\lambda) = - \int_{-1}^1 \frac{1}{(M(y) - \lambda)} \; \frac{M''(y)}{M'(y)^2}
\; dy  - \Big[ \frac{1}{(M(y) - \lambda)\, M'(y)}\Big]_{-1}^{1}.
$$
That is to say, for $\lambda = \lambda_R + i \lambda_I$,
$$
\left| \begin{array}{l}
\displaystyle {\cal R}e \;F(\lambda) = - \left\{ \int_{-1}^1 \frac{M(y) - \lambda_R}{|M(y) - \lambda|^2}
  \; \frac{M''(y)}{M'(y)^2} \; dy
+ \Big[ \frac{M(y) - \lambda_R}{|M(y) - \lambda)|^2 \, M'(y)}\Big]_{-1}^{1}
\right\},
\\[18pt]
\displaystyle {\cal I}m \;F(\lambda) = - \lambda_I \left\{ \int_{-1}^1 \frac{1}{|M(y) - \lambda|^2}
  \; \frac{M''(y)}{M'(y)^2} \; dy
+ \Big[ \frac{1}{|M(y) - \lambda)|^2 \, M'(y)}\Big]_{-1}^{1} \right\}.
\end{array} \right.$$
If $F(\lambda) = 2$ with $\lambda_I \neq 0$, we have in particular
\beq \label{ident}
\int_{-1}^1 \frac{1}{|M(y) - \lambda|^2}
  \; \frac{M''(y)}{M'(y)^2} \; dy
+ \Big[ \frac{1}{|M(y) - \lambda)|^2 \, M'(y)}\Big]_{-1}^{1} = 0.
\ene
Substituting this in the right hand side of ${\cal R}e \;F(\lambda)$
(note that the above term is the multiplicative factor of
$\lambda_R$), we get (since $M(-1) =0$ and setting $M'_{\pm}
=M'(\pm1)$ )
$$
 {\cal R}e \;F(\lambda) = - \int_{-1}^1 \frac{M(y)}{|M(y) - \lambda|^2}
  \; \frac{M''(y)}{M'(y)^2} \; dy -
 \frac{M_{+}}{|M_{+} - \lambda|^2 \, M'_{+}}.
$$
Next, we claim that ${\cal R}e \;F(\lambda) < 0 $ so that $,F(\lambda)=2$ is
impossible. We distinguish two cases:
\begin{itemize}
\item[(i)] $M''(y) > 0$. In this case, ${\cal R}e \;F(\lambda)$ is clearly the sum of
    two negative terms.
\item[(ii)] $M''(y) < 0$. In this case, we can write
$$
\left| \begin{array}{ll}
\displaystyle - \int_{-1}^1 \frac{M(y)}{|M(y) - \lambda|^2}
  \; \frac{M''(y)}{M'(y)^2} \; dy & \displaystyle
\leq  - \int_{-1}^1 \frac{M_+}{|M(y) -
    \lambda|^2}
  \; \frac{M''(y)}{M'(y)^2} \\[12pt]
& \displaystyle = \Big[ \frac{M_+}{|M(y) - \lambda)|^2 \, M'(y)}\Big]_{-1}^{1}.
\end{array} \right.
$$
The second inequality being deduced from (\ref{ident}). Then,
$$
 {\cal R}e \;F(\lambda) \leq  \Big[ \frac{M_+}{|M(y) - \lambda)|^2 \,
   M'(y)}\Big]_{-1}^{1} -
 \frac{M_{+}}{|M_{+} - \lambda|^2 \, M'_{+}} = -\frac{M_+}{|\lambda|^2 \,
   M'_-} <0.
$$
\end{itemize}
This achieves the proof.

\bull \\[12pt]

Next, we wish to study $F(\lambda)$ when $\lambda$
approaches the segment $[M_-,M_+]$, assuming that $M \in {\cal C}_s^2$. The idea is to use a well known result
from the theory of Cauchy integrals (see \cite{ku}, \cite{mu}, and
\cite{pri}). Let us first give some technical definitions.
\begin{definition} \label{Holder}
Defining the complex open halfspaces as
$\CE^\pm = \{ \lambda \in \CE \, / \, \pm {\cal I}m \; \lambda > 0 \}$
we say that $h_\pm : \overline{\CE^\pm} \longrightarrow \CE$ is locally Hölder
continuous in $\overline{\CE^\pm}$ with exponent $\gamma \in \; ]0,1[$ if
\beq \label{Hol1}
\forall \; R >0, \quad \exists \; C_R>0 \quad / \quad  \forall \; (\lambda_1,
\lambda_2) \in {\cal B}_R^\pm, \quad |h_\pm(\lambda_1) - h_\pm(\lambda_2)| \leq C_R \; |\lambda_1
-\lambda_2|^{\gamma}
\ene
where $  {\cal B}_R^\pm = \{ \lambda \in \overline{\CE^\pm} \, / \, |\lambda| \leq
R$\}. If the constant  $C_R$ can be taken independent of $R$ we say that $h_\pm$ is H\"older continuous in $\overline{\CE^\pm}$.\\[12pt]
In the same way we say that $h_\pm : \overline{\CE^\pm} \setminus \{M_-,M_+\}  \longrightarrow \CE$
is locally Hölder
continuous in $\overline{\CE^\pm} \setminus \{M_-,M_+\}$ with exponent $\gamma
\in \; ]0,1[$ if
\beq \label{Hol1b}
\forall \; R, \delta >0, \quad \exists \; C_{R,\delta} >0 \quad / \quad
\forall \; (\lambda_1, \lambda_2) \in {\cal B}_{R,\delta}^\pm, \quad |h_\pm(\lambda_1) -
h_\pm(\lambda_2)| \leq C_{R,\delta} \, |\lambda_1 -\lambda_2|^{\gamma}
\ene
where $  {\cal B}_{R,\delta}^\pm = \{ \lambda \in  {\cal B}_R^\pm \, / \, |\lambda-M_\pm| \geq
\delta \}$.\\[12pt]
\end{definition}
We quote below a classical result in Cauchy integrals that we use.
\begin{prop} \label{propo} (Plemelj-Privalov Theorem)
Let $g(z): [M_-, M_+] \rightarrow \CE$ be Hölder continuous with
  exponent $\gamma \in \; ]0,1[$ in $[M_-,M_+]$. Define the function
$$
S(\lambda )= \int_{M_-}^{M_+}\,\, g(z) \,
(z-\lambda)^{-1}\, dz , \quad \lambda \in \CE \setminus [M_-,M_+].
$$
Then, for any $\lambda \in \CE \setminus ]M_-,M_+[$
\beq \label{res}
\lim_{\ep \downarrow 0}\, S(\lambda \pm i \ep) = S_\pm(\lambda) :=
P. V. \,  \int_{M_-}^{M_+}\,\, g(z) \,
(z-\lambda)^{-1}\, dz \,\,
 \pm i\pi g(\lambda),
\ene
where  $P. V.$ stands for the Cauchy principal value of the integral. The
function
\beq
S_\pm(\lambda):= S(\lambda), \quad\mbox{if } \lambda \in \overline{\CE^\pm} \setminus
  [M_-,M_+] \quad \mbox{and}
\quad S_\pm(\lambda), \quad \mbox{if } \lambda\in  \; ]M_-,M_+[,
\label{3.33}
\ene
is analytic in $\CE^\pm$ and locally H\"older
continuous with exponent $ \gamma$ in $\overline{\CE^\pm} \setminus
\{M_-,M_+\} $. Moreover, if $g(M_+) = g(M_-)= 0$ , $S_\pm(\lambda)$
can be continuously extended to $\overline{\CE^\pm}$ into a function
which is locally H\"older continuous in $\overline{\CE^\pm}$.
Conversely, if $g(M_\pm) \neq 0$, $S_\pm(\lambda)$ blows up as a
logarithm when $\lambda$ tends to $M_\pm$.
\end{prop}
\noindent In what follows, we assume that $M \in  {\cal C}^{3,
  \gamma}_s$. For the sequel, it will be useful to introduce $\mu(z)$ the
inverse function to
$M(y)$,
\beq
\mu(z)=y, \quad \hbox{\rm for}\, z=M(y), \quad  y \in [-1,1].
\label{3.28}
\ene
Changing  the variable of integration to $z= M(y)$, we have that,
\beq
F(\lambda)= \int_{M_-}^{M_+}\, \mu'(z) \, (z-\lambda)^{-2}\, dz, \quad \lambda \in \CE \setminus [M_-,M_+].
\label{3.29}
\ene
In order to apply Proposition \ref{propo}, we integrate by parts in
(\ref{3.29}) and obtain that,
\beq
F(\lambda)=- \mu'(z)\, (z-\lambda)^{-1}\Big|^{M_+}_{M_-}+ \int_{M_-}^{M_+}\,
\mu''(z) \, (z-\lambda)^{-1}\, dz.
\label{3.30}
\ene
Applying Proposition \ref{propo} with $g = \mu''$, we get\\
\beq \label{limF}
\left\{ \begin{array}{l}
\lim_{\ep \downarrow 0}\, F(\lambda \pm i \ep) = F_\pm(\lambda), \quad \forall
\; \lambda \in ]M_-,M_+[,
\\[12pt]
\displaystyle F_\pm(\lambda):=  - \mu'(z)\, (z-\lambda)^{-1}\Big|^{M_+}_{M_-} +
P. V. \,  \int_{M_-}^{M_+}\,\, \mu''(z) \,
(z-\lambda)^{-1}\, dz \,\,
 \pm i\pi \mu''(\lambda),
\end{array} \right.
\ene
and we can define as in Proposition \ref{propo}, formula (\ref{3.33}),
$F_\pm(\lambda)$ as a function that is analytic in $\CE^\pm$ and locally H\"older
continuous with exponent  $ \gamma$ in
$\overline{\CE^\pm} \setminus \{M_-,M_+\} $.
Note that for $\lambda \in \; ]M_-,M_+[, {\cal I}m \; F_\pm(\lambda) = \pm \pi \,
\mu''(\lambda) \neq 0$,  in particular $F_\pm(\lambda) - 2 \neq 0$,  which allows us to state the following result.

\begin{lemma}
\label{le3.1b}
Assume that $M \in {\cal C}^{2, \gamma}_s$. Then, for any $\lambda \in \; ]M_-,M_+[$
\beq \label{resb}
\lim_{\ep \downarrow 0}\, N(\lambda \pm i \ep) = \big(2- F_{\pm}(\lambda)\big)^{-1}.
\ene
Moreover, the function
\beq
N_\pm(\lambda):= \left\{ \begin{array}{ll} N(\lambda), & \lambda \in \CE^\pm
= \{ z \in \CE \, / \, \pm {\cal I}m \, z > 0 \},
\\\\ \big( F_{\pm}(\lambda) -2
\big)^{-1}, & \lambda  \in  \; ]M_-,M_+[,
\end{array}\right.
\label{3.33b}
\ene
is extended by continuity to $\overline{\CE^\pm}$, with $N_\pm(M)= 0$ for $M
= M_+$ or $M_-$, as a
function  analytic in $\CE^\pm$ and  H\"older continuous in $\overline{\CE^\pm}$ with exponent $\gamma$.
Moreover,
\beq
\begin{array}{l}
\forall \; \lambda \in [M_-,M_+], \quad N_-(\lambda) = \overline{N_+(\lambda)},
\quad \mbox{i. e. } N_\pm(\lambda)= {\cal R}e\; N(\lambda) \pm i \,{\cal I}m\; N(\lambda),
\end{array}
\label{3.36}
\ene
where ${\cal R}e\; N$ is the common value of ${\cal}Re \;N_{+}$ and ${\cal R}e\; N_{-}$ and
${\cal I}m \;N := {\cal I}m \;N_+ = - {\cal I}m \;N_-$.
\end{lemma}
\begin{remark} One can easily obtain an explicit expression of ${\cal R}e\; N$ and ${\cal I}m
 \; N$ from (\ref{limF}) and (\ref{3.33}). For instance, one has

\beq {\cal I}m\; N(\lambda)=  \pi \, |N_\pm(\lambda)|^2  \;
\mu''(\lambda), \quad \lambda \in [M_-,M_+].
\label{3.37}
\ene Such
expressions are, in particular, useful for numerical computations
(see \cite{jjw}).
\end{remark}

\noindent {\it Proof:} The continuous extension by
$N_\pm(M_\pm)= 0$ is valid  by (\ref{3.30}). The local H\"older continuity of
$N_\pm$  in $\overline{\CE^\pm} \setminus \{M_-,M_+\} $ is inherited from the
same property for $F_\pm$. The H\"older continuity up to $M_\pm$ remains to be
clarified. We consider the case of $M_+$ ($M_-$ is treated analogously) and
introduce the function
$$
J^+(\lambda) = (M_+ - \lambda) \, \big( 2 - F_\pm(\lambda) \big), \quad \mbox{
  such
  that } N_\pm(\lambda) = (M_+ - \lambda) \, J^+(\lambda)^{-1}.
$$
We can conclude if we show that $J^+$ is Hölder continuous in a neighborhood
of $M_+$ and does not vanish at $M=M_+$. According to (\ref{3.30}),
$$
J^+(\lambda) =  \mu'(M_+) + {(M_+-\lambda)} \; \Big( 2 -  \frac{\mu'(M_-)}{M_--\lambda} -
 \int_{M_-}^{M_+}\,\, \frac{\mu''(z)}{z-\lambda} \, dz \Big) ,
$$
that we rewrite as
\beq \label{expJ0+}
\left| \begin{array}{lll}
J^+(\lambda) =  \mu'(M_+) & + & \ds {(M_+-\lambda)} \; \Big( 2 -
\frac{\mu'(M_-)}{M_--\lambda} - \mu''(M_+) \log \left[
 \frac{M_+-\lambda}{M_--\lambda} \right]\Big)\\[12pt]& - &\ds {(M_+-\lambda)}
\;
 \int_{M_-}^{M_+}\,\, \frac{\mu''(z) - \mu''(M_+)}{z-\lambda} \, dz.
\end{array} \right.
\ene
Consequently, since
\begin{itemize}
\item $x \mapsto x \, \log x$ is Hölder continuous with any exponent in $(0,1)$,
\item by Proposition \ref{propo} applied with $g(z) = \mu''(z) - \mu''(M_+)$
  (that satisfies $g(M_+)= 0$),
$$
\lambda \rightarrow \int_{M_-}^{M_+}\,\, \frac{\mu''(z) -
  \mu''(M_+)}{z-\lambda} \, dz \quad \mbox{ is Hölder continuous with
  exponent } \gamma,
$$
\end{itemize}
we deduce from (\ref{expJ0+}) that $J_+$ is Hölder continuous with exponent $
\gamma$ in a neighborhood of $M_+$. Finally, $J^+(M_+) = \mu'(M_+) \neq 0$, since
$M \in {\cal C}^{2,\gamma}_s$.

\bull \\[12pt]
We shall need an analogous result for $N'(\lambda)$, the derivative of
$N(\lambda)$. For this, we assume that $M \in {\cal C}^{3, \gamma}_s$
and remark that for $\lambda \in \CE \setminus [M_-, M_+]$
\beq
\left| \begin{array}{lll}
F'(\lambda)& = & \ds - \int_{M_-}^{M_+}\, \mu'(z) \,\frac{d}{dz} \left[
  (z-\lambda)^{-2} \right] \, dz \\[12pt]
& = & \ds - \mu'(z) \,
  (z-\lambda)^{-2} \Big|^{M_+}_{M_-} + \int_{M_-}^{M_+}\,
\mu''(z) \, (z-\lambda)^{-2}\, dz.
\\[12pt]
& = & \ds - \mu'(z) \,
  (z-\lambda)^{-2} \Big|^{M_+}_{M_-}- \mu''(z)\, (z-\lambda)^{-1}\Big|^{M_+}_{M_-}+ \int_{M_-}^{M_+}\,
\mu'''(z) \, (z-\lambda)^{-1}\, dz.
\end{array} \right.
\label{Fprime}
\ene
Applying Proposition \ref{propo} as for $F(\lambda)$, we obtain
\beq \label{limFprime}
\left\{ \begin{array}{l}
\lim_{\ep \downarrow 0}\, F'(\lambda \pm i \ep) = F'_\pm(\lambda), \quad \forall
\; \lambda \in ]M_-,M_+[,
\\[12pt]
\begin{array}{lll}
\displaystyle F'_\pm(\lambda)& := &  - \mu'(z)\,
(z-\lambda)^{-2}\Big|^{M_+}_{M_-}
- \mu''(z)\, (z-\lambda)^{-1}\Big|^{M_+}_{M_-} \\[12pt]
& + & \ds P. V. \,  \int_{M_-}^{M_+}\,\,
\mu'''(z) \,
(z-\lambda)^{-1}\, dz \,\,
 \pm i\pi \mu'''(\lambda),
\end{array}
\end{array} \right.
\ene
and we can define again as in Proposition \ref{propo}, formula (\ref{3.33}),
$F'_\pm(\lambda)$ as a function that is analytic in $\CE^\pm$ and locally H\"older
continuous with exponent  $ \gamma$ in
$\overline{\CE^\pm} \setminus \{M_-,M_+\} $. Next we state for $N'$ the
equivalent result of Lemma \ref{le3.1} for $N$.
\begin{lemma}
\label{le3.1prime}
Assume that $M \in {\cal C}^{3, \gamma}_s$. Then, for any $\lambda \in \; ]M_-,M_+[$
\beq \label{resc}
\lim_{\ep \downarrow 0}\, N'(\lambda \pm i \ep) = - F'_\pm(\lambda) \; \big(2- F_{\pm}(\lambda)
\big)^{-2}.
\ene
Moreover, the function
\beq
N'_\pm(\lambda):= \left\{ \begin{array}{ll} N'(\lambda), & \lambda \in \CE^\pm
= \{ z \in \CE \, / \, \pm {\cal I}m \; z > 0 \},
\\\\ - F'_\pm(\lambda) \; \big( 2-F_{\pm}(\lambda)
\big)^{-2}, & \lambda  \in  \; ]M_-,M_+[,
\end{array}\right.
\label{3.33c}
\ene
is extended by continuity to $\overline{\CE^\pm}$ with
\beq \label{extNprime}
\ds N'_\pm(M)=
- \frac{\mu''(M)}{\mu'(M)^2} \quad \mbox{for } M = M_+ \mbox{ or } M_-.
\ene
Then, $N'_\pm(\lambda)$ is
  analytic in $\CE^\pm$ and  H\"older continuous in
$\overline{\CE^\pm}$ with exponent $\gamma$. Moreover,
\beq
\begin{array}{l}
\forall \; \lambda \in [M_-,M_+], \quad N'_-(\lambda) = \overline{N'_+(\lambda)},
\quad \mbox{i. e. } N'_\pm(\lambda)= {\cal R}e\; N'(\lambda) \pm \,i\, {\cal I}m\; N'(\lambda),
\end{array}
\label{3.36b}
\ene
where ${\cal R}e\; N'$ is the common value of ${\cal R}e \;N'_{+}$ and ${\cal R}e \;N'_{-}$ and
${\cal I}m\; N' = {\cal I}m \;N'_+ = - {\cal I}m \;N'_-$.
\end{lemma}
\noindent {\it Proof:} It is very similar to the proof of Lemma
\ref{le3.1}. From formula (\ref{Fprime}), using the same trick used for proving
the Hölder continuity of  $J^+(\lambda)$ (see the proof of Lemma
\ref{le3.1}), we deduce that:
$$
F'_\pm(\lambda) = (M_+-\lambda)^{-2} \big[ \,- \mu''(M_+) + h(\lambda) \big],
\quad h \in  {\cal C}^{0, \gamma}, \quad h(0) = 0.
$$
Therefore, since $N_\pm(\lambda)^2=  (M_+-\lambda)^{2}\; J^+(\lambda)^{-2}$,
$$
N'_\pm(\lambda)= - F'_\pm(\lambda) \; N_\pm(\lambda)^2 =  \big[ \, -\mu''(M_+) +
h(\lambda) \big] \; J^+(\lambda)^{-2}.
$$
It is then easy to conclude.

\bull \\[12pt]
\begin{remark} Again  we can easily obtain an explicit expression of ${\cal R}e\; N'$ and ${\cal I}m\;
  N'$ from (\ref{limFprime}) and (\ref{3.33b}) (see \cite{jjw}).
\end{remark}
With Lemmata \ref{le3.1} and \ref{le3.1prime}, we now have all the needed
information for studying the integrals $I_\ell(kt,y), \ell = 0,1$. We first
obtain a new expression for $I_{\ell,c}(kt,y)$:
\begin{lemma} \label{I0I1}
Assume that $M \in {\cal C}^{2,\gamma}_s$, then
\beq \label{expI0}
 \left| \begin{array}{lll}I_{0,c}(kt,y) &:= & \displaystyle
    P. V. \int_{M_-}^{M_+}\, {e^{-ik\lambda t}}
\; \frac{\big[\,N_+(\lambda) - N_-(\lambda)\big] }{\big(\lambda-M(y)\big)} \; d
\lambda \\[18pt]
& - & \displaystyle i \pi \big[ \,N_+\big(M(y)\big) + N_-\big(M(y)\big) \, \big] \; {e^{-ik M(y) t}}.
\end{array} \right.
\ene If moreover, $M \in {\cal C}^{3,\gamma}_s$, then
\beq
\label{expI1} \left| \begin{array}{lll} \displaystyle I_{1,c}(kt,y)
& := & \displaystyle - ikt \; I_{0,c}(kt,y) + P. V.
\int_{M_-}^{M_+}\, {e^{-ik\lambda t}} \; \frac{\big[ \,N'_+(\lambda)
- N'_-(\lambda) \, \big] }{\big(\lambda-M(y)\big)} \; d
\lambda  \\[18pt]
& - & \displaystyle i \pi \big[ \, N'_+\big(M(y)\big) + N'_-\big(M(y)\big)\, \big]
\; {e^{-ik M(y) t}} .
\end{array} \right.
\ene
\end{lemma}
\noindent {\it Proof:} Let us start from the formula (\ref{3.21}) for
$I_{0,c}(kt,y)$ that we can rewrite:
$$I_{0,c}(kt,y) :=  \sum_{\pm} \int_{\gamma_{\delta}^\pm}
\; \frac{N_{\pm,kt}(\lambda)}{\big(\lambda-M(y)\big)} \; d \lambda
+\sum_{\pm} (\pm 1) \cdot \int_{M_-}^{M_+}
\; \frac{N_{kt}(\lambda\pm i \delta)}{\big(\lambda\pm i \delta-M(y)\big)}
\; d \lambda $$
where $N_{\pm, kt}(\lambda) := N_\pm(\lambda) \; e^{-ikt\lambda}$. Since $N(M_\pm)=0$,
$$ \displaystyle
\lim_{\delta \rightarrow 0} \int_{\gamma_{\delta}^\pm}
\; \frac{N_{kt}(\lambda)}{\big(\lambda-M(y)\big)} \; d \lambda = 0,
$$
so that
$$
I_{0,c}(kt,y) = \lim_{\delta \rightarrow 0}\int_{M_-}^{M_+}
\; \frac{N_{+,kt}(\lambda + i \delta)}{\big(\lambda+ i \delta-M(y)\big)}\; d \lambda  -
\lim_{\delta \rightarrow 0}\int_{M_-}^{M_+}
\; \frac{N_{-,kt}(\lambda - i \delta)}{\big(\lambda- i \delta-M(y)\big)}\; d \lambda,
$$
that we rewrite as
$$
I_{0,c}(kt,y) = \Big[ \, I_{0,c}^{(1,+)}(kt,y) - I_{0,c}^{(1,-)}(kt,y) \, \Big]
+ \Big[ \,I_{0,c}^{(2,+)}(kt,y) - I_{0,c}^{(2,-)}(kt,y)\, \Big],
$$
where
$$
\left| \begin{array}{l}
\displaystyle I_{0,c}^{(1,\pm)}(kt,y) = \lim_{\delta \rightarrow 0}\int_{M_-}^{M_+} \;
\frac{N_{\pm, kt}(\lambda\pm i \delta) - N_{\pm,kt}(\lambda)}{\big(\lambda\pm i
  \delta-M(y)\big)}, \\[18pt]
\displaystyle I_{0,c}^{(2,\pm)}(kt,y) = \lim_{\delta \rightarrow 0}    \int_{M_-}^{M_+}
\; \frac{N_{\pm,kt}(\lambda)}{\big(\lambda\pm i \delta-M(y)\big)}.
\end{array} \right.
$$
The Hölder continuity of $N_{\pm,kt}(\lambda)$ (see Lemma \ref{le3.1}) implies
$I_{0,c}^{(1,\pm)}(kt,y) = 0$. Then, it is sufficient to apply again
Proposition \ref{propo} with $g = -N(\lambda) \,e^{-ikt} $ to get the result.\\[12pt]
For $I_{1,c}(kt,y)$, we first integrate by parts in (\ref{3.21}) to obtain
$$
 I_{1,c}(kt,y) :=  \int_{{\cal S}_{\delta}}\,
\; \frac{N_{kt}'(\lambda)}{\big(\lambda-M(y)\big)} \; d \lambda.
$$
We then proceed as above using $N_{\pm,kt}'(\lambda) = \big( N_\pm'(\lambda) - ikt
N_\pm(\lambda) \big) \; e^{-ikt\lambda}$ and Lemma \ref{le3.1prime}. The details
are left to the reader.

\bull\\[12pt]
We are now in position to derive our estimates for $I_{\ell}(kt,y), \ell =0,1$.

\begin{lemma} \label{estI1I2} Assume that $M \in {\cal C}^{\ell +1}_s, \ell =
  1,2$. Then, there exists a constant $I_{\ell}^* > 0$, depending only on $\ell$ and
  $M$, such that one has the uniform
  estimates:
\beq \label{estI0I1}
 \forall \; k \in \ere, \quad \forall \; t > 0, \quad
\|I_{\ell}(kt,\cdot)\|_{L^{\infty}([-1,1])}  \leq I_{\ell}^* \; (1 + \ell \;
|kt|^\ell), \quad \ell = 0,1.
\ene
\end{lemma}
\noindent {\it Proof:}
Remarking that, thanks to (\ref{3.36}),
$
(N_+ + N_-)(\lambda)= 2 {\cal R}e\; N(\lambda)$ and setting

\beq \label{DiffN}
\Delta N (\lambda):=(N_+ - N_-)(\lambda)= 2i ({\cal I}m\; N(\lambda)
\equiv 2 \, i \,
\pi \, |N_\pm(\lambda)|^2  \mu''(\lambda),
\ene
by (\ref{3.36})
and (\ref{3.37}). From (\ref{expI0}) we deduce that,
\beq \label{Proof1} I_{0,c}(kt,y) \leq
 | \, P. V. \int_{M_-}^{M_+}\, {e^{-ik\lambda t}}
\; \frac{\Delta N(\lambda)}{\big(\lambda-M(y)\big)} \; d
\lambda \, |  + 2 \pi \; {\cal R}e\; N\big(M(y)\big)|.
\ene
By Lemma \ref{le3.1}  we know that $\Delta N(\lambda)$ is Hölder continuous
with exponent $\gamma$. Since $\Delta N(\lambda)=0$, for $ \lambda \leq M_-$ and for $\lambda \geq M_+$,
we can write,
denoting  $L = M_+ - M_-$ (so that $M(y) -L \leq M_-$ and $M(y)+L \geq M_+$),
\beq \label{Proof2}
\left| \begin{array}{ll}
\displaystyle P. V. \int_{M_-}^{M_+}\, {e^{-ik\lambda t}}
\; \frac{\Delta N(\lambda)}{\big(\lambda-M(y)\big)} \; d
\lambda   \displaystyle = P. V. \int_{M(y) -L}^{M(y) + L}\, {e^{-ik\lambda t}}
\; \frac{\Delta N(\lambda)}{\big(\lambda-M(y)\big)} \; d
\lambda \\[18pt]
\quad \quad \quad \quad \quad \quad \quad \quad
 \displaystyle = \;  e^{-ikM(y) t} \; P. V. \int_{-L}^{L}\, {e^{-ik \nu t}}
\; \frac{\Delta N\big(\nu + M(y)\big)}{\nu} \; d
\nu.
\end{array} \right.
\ene
Next, we write
\beq \label{Proof3}
\left| \begin{array}{ll}
\displaystyle P. V. \int_{-L}^{L}\, {e^{-ik \nu t}}
\; \frac{\Delta N\big(\nu + M(y)\big)}{\nu} \; d
\nu & \displaystyle = \int_{-L}^{L}\, {e^{-ik \nu t}}
\; \frac{\Delta N\big(\nu + M(y)\big) - \Delta N\big(M(y)\big)}{\nu}
\; d \nu \\[18pt]
& \displaystyle + \; {\Delta N}\big(M(y)\big) \; \Big( P.V. \int_{-L}^{L}\,
\frac{e^{-ik \nu t}}{\nu} \; d \nu \Big). \end{array} \right.
\ene
Finally, we remark that
$$
 P.V. \int_{-L}^{L}\,
\frac{e^{-ik \nu t}}{\nu} \; d \nu = -i\int_{-L}^{L}\,
\frac{\sin({k \nu t})}{\nu} \; d \nu = -i \int_{-Lkt}^{Lkt}\,
\frac{\sin(\xi)}{\xi} \; d \xi
$$
is bounded (in modulus) since
$$
\displaystyle \Big|\int_{- \infty}^{+ \infty}\,
\frac{\sin(\xi)}{\xi} \; d \xi\Big| = \lim_{A \rightarrow + \infty} \Big|\int_{-A}^{A}\,
\frac{\sin(\xi)}{\xi} \; d \xi\Big| < + \infty.
$$
Finally, if we set
\beq \label{Proof4}
K:= \sup_{A>0} \Big|\int_{-A}^{A}\,
\frac{\sin(\xi)}{\xi} \; d \xi\Big| \quad \mbox{and} \quad
|\Delta {N}|_{\gamma} := \sup_{(x,y)\in \ere^2}
\frac{|\Delta N(x) -\Delta N(y)|}{|x-y|^\gamma} \; ,
\ene
we deduce from (\ref{Proof2}) and (\ref{Proof3}) that ($\|\cdot\|_{L^\infty}$
meaning $\|\cdot\|_{L^\infty(-M_-,M_+)}$):
\beq \label{Proof5}
\displaystyle \Big| P. V. \int_{-L}^{L}\, {e^{-ik \nu t}}
\; \frac{\Delta N\big(\nu + M(y)\big)}{\nu} \; d
\nu \Big|  \displaystyle \leq
\Big(\int_{-L}^L \frac{d \nu}{|\nu|^{1-\gamma}}\Big) \; |\Delta N|_{\gamma} + K
\; \|\Delta {N}\|_{L^\infty},
\ene
which, using (\ref{Proof1}), gives:
\beq \label{Proof6}
|I_{0,c}(kt,y)| \leq I_{0,c}^* := 2 \,   \frac{|M_+ - M_-|^{\gamma}}{\gamma} \; |\Delta {N}|_{\gamma} + K
\; \|\Delta {N}\|_{L^\infty} + 2 \pi \|\Re N\|_{L^\infty} .
\ene
Finally, since (see (\ref{3.19}, \ref{3.21})),
$$
|I_{0}(kt,y)| \leq 2\pi \sum_{\pm} \frac{ |\mbox{Res}\,
  N(\lambda_\pm)|}{|\lambda_\pm-M(y)|} + |I_{0,c}(kt,y)|
$$
we obtain the inequality  (\ref{estI0I1}) for $\ell = 0$ with
$$
I_0^* =  I_{0,c}^* + 2\pi \sum_{\pm} |\mbox{Res}\,
  N(\lambda_\pm)| \; \| \frac{1}{\lambda_\pm-M(\cdot)}\|_{L^\infty}.
$$
The inequality  (\ref{estI0I1}) for $\ell = 1$ is derived analogously using (\ref{expI1}) and
Lemma \ref{le3.1prime}. This time, instead of $\Delta N$, we have to work with
$\Delta N' : = N_+' - N_-'$ which fortunately satisfies $\Delta N'(\lambda)=0$ for $ \lambda \in (-\infty, M_-] \cup [M_+, \infty)$,
thanks to (\ref{extNprime}). Note
that the additional factor $|kt|$ comes from the first term in the right-hand side  of (\ref{expI1}).

\bull \\[12pt]
We now go back to the estimates of $\hat{\bf a}_0$ and $\hat{\bf a}_1$. Using
the fact that
$
\ds |{\bf a}(uv)| \leq \|u\|_{L^2_y} \, \|v\|_{L^2_y}
$
(Cauchy-Schwartz) and $\|u\|_{L^2_y} \leq \sqrt{2} \, \|u\|_{L^\infty_y}$, we
deduce from (\ref{ha0}) and Lemma \ref{estI1I2} that for $n=0,1,\cdots$,
\beq \label{estha0}
|k|^{1+n} \; |\hat{\bf a}_0(k,t)| \leq \frac{\sqrt{2}}{\pi} \; \Big( I_0^* + I_1^*
\; \|M\|_{L^\infty} \big(1 + |kt| \big)\Big) \; |k|^{1+n} \; \|\hau^0(k, \cdot)\|_{L^2_y}.
\ene
In the same way, from (\ref{ha1}) and again Lemma \ref{estI1I2}, we obtain
\beq \label{estha1}
|k|^{1+n} \; |\hat{\bf a}_1(k,t)| \leq \frac{\sqrt{2}}{\pi} \; I_1^*
\; \big(1 + |kt| \big)\; |k|^{n} \; \|\hau^1(k, \cdot)\|_{L^2_y}, \quad n=0,1,\cdots.
\ene
Thus, by Plancherel's theorem, we see that, $C$ denoting a constant depending
on $\|M\|_{L^\infty}, I_0^*$ and $I_1^*$ (and thus on $M$ only),
\beq \label{esthau}
\left| \begin{array}{ll}
\ds \| \frac{\partial^{1+n} {\bf a}(u)}{\partial x^{1+n}}
(\cdot,t)\|_{L^2_x} \; \leq \; C&\ds \Big( \; \|\frac{\partial^{1+n} u^0}{\partial
  x^{1+n}}\|_{L^2_y(L^2_x)} +  \|\frac{\partial^{n} u^1}{\partial
  x^{n}}\|_{L^2_y(L^2_x)}\\[12pt]
& \; \; \; \ds + \; \; t \; \Big[ \, \|\frac{\partial^{2+n} u^0}{\partial
  x^{2+n}}\|_{L^2_y(L^2_x)} + \|\frac{\partial^{1+n} u^1}{\partial
  x^{1+n}}\|_{L^2_y(L^2_x)} \, \Big] \; \Big),
\end{array} \right.
\ene
for $ n=0,1,\cdots$.

From (\ref{3.1}), or equivalently (\ref{3.3}), one easily
deduces that for each $y \in ]-1,1[$ and $t \geq 0$:
\beq
\label{estu1}
\begin{array}{l}\ds
u(x,y,t)= u^0(x-t M(y),y)+ t \left(u^1(x-M(y)t,y)+ M(y)
  \frac{\partial}{\partial x}u^0(x-M(y)t, y)\right)\\\\ \hspace{1.5cm}+
\int_0^t\, dz_1 \, \int_0^{z_1}\, dz_2\,\frac{\partial^{2} {\bf a}(u)}{\partial x^2}(x- t M(y)+ z_2 M(y), z_2),
\end{array}
\ene
for all $ t >0,  (x,y)
\in \ere \times [-1,1]$. Taking the $L^2$-norm in $y$ and using (\ref{esthau}), we easily obtain (with
another constant $C$ depending only on $M$)

\beq \label{estu2}
\left| \begin{array}{l} \ds\|\frac{\partial^n u(x,y,t)}{\partial x^n}\|_{L^2_y(L^2_x)} \; \leq  \ds
    \|\frac{\partial^n u^0}{\partial x^n}(\cdot,y)\|_{L^2_y(L^2_x)}  +  \ds t \; \Big( \;
\|\frac{\partial^n u^1(\cdot,y)}{\partial x^n}\|_{L^2_y(L^2_x)} +\\[12pt]
\ds\|M\|_{L^\infty} \; \|\frac{\partial^{1+n}
  u^0}{\partial x^{1+n}} (\cdot,y)\|_{L^2_y(L^2_x)} \; \Big)
 \ds +  C \; \ds t^2 \; \Big( \, \|\frac{\partial^{2+n} u^0}{\partial
  x^{2+n}}\|_{L^2_y(L^2_x)} +  \|\frac{\partial^{1+n} u^1}{\partial
  x^{1+n}}\|_{L^2_y(L^2_x)} \, \Big)
 \\[12pt]
 \ds +  \ds C \; t^3 \; \Big( \, \|\frac{\partial^{3+n} u^0}{\partial
  x^{3+n}}\|_{L^2_y(L^2_x)} +  \|\frac{\partial^{2+n} u_1}{\partial
  x^{2+n}}\|_{L^2_y(L^2_x)} \, \Big), \quad n=0,1,\cdots.
\end{array} \right.
\ene
Concerning $\ds \frac{\partial u}{\partial t}$ we observe that taking the time derivative amounts to the following: in the first equation in
(\ref{3.21}) to multiplication by $ -ik \lambda_\pm$, and in equations (\ref{expI0}, \ref{expI1}) to multiplication by $ -ik \lambda, \lambda \in
[M_-,M_+]$ and by $-ik M(y)$. We easily see that arguing as above one obtains
estimates for $\ds |\frac{\partial \hat{\bf a}_0}{\partial t} (k,t)|$
and for $\ds |\frac{\partial \hat{\bf a}_1}{\partial t} (k,t)|$ which are
similar to those for $\ds |\hat{\bf a}_0 (k,t)|$
and $\ds |\hat{\bf a}_1(k,t)|$ up to an extra
power of $|k|$. As a consequence, it is easy to obtain the following estimate
for the mean value of   ${\bf a}(u)$:
\beq \label{estderhau}
\left| \begin{array}{l}
\ds \| \frac{\partial^{1+n} {\bf a}(u)}{\partial t \partial x^n}
(x,t)\|_{L^2_x} \; \leq \; C\ds \Big( \; \|\frac{\partial^{1+n} u^0}{\partial
  x^{1+n}}\|_{L^2_y(L^2_x)} +  \|\frac{\partial^{n} u^1}{\partial
  x^{n}}\|_{L^2_y(L^2_x)}
 \; \; \; \ds + \\[12pt] t \Big[ \, \|\frac{\partial^{2+n} u^0}{\partial
  x^{2+n}}\|_{L^2_y(L^2_x)} + \|\frac{\partial^{1+n} u^1}{\partial
  x^{1+n}}\|_{L^2_y(L^2_x)} \, \Big] \; \Big), \quad n= 0,1,\cdots.
\end{array} \right.
\ene
Furthermore, taking the time derivative in both sides of (\ref{estu1}), we obtain that,
\beq \label{estu3b}
\left| \begin{array}{l}
\ds\|\frac{\partial^{1+n} u}{\partial t \partial x^n}(x,y,t)\|_{L^2_y(L^2_x)} \; \leq  \ds
    \|(1+M(y))\frac{\partial^{n+1}  u^0}{\partial x^{n+1}}\|_{L^2_y(L^2_x)}  +
    \|\frac{\partial^{n}  u^1}{\partial x^{n}}\|_{L^2_y(L^2_x)}  +\\[12pt] C t \; \Big(
    \; \|\frac{\partial^{2+n} u^0}{\partial x^{2+n}}\|_{L^2_y(L^2_x)}  +   \|\frac{\partial^{n+1}  u^1}{\partial x^{n+1}}\|_{L^2_y(L^2_x)}
\Big)+
\ds C \; t^2 \; \Big( \, \|\frac{\partial^{3+n} u^0}{\partial
  x^{3+n}}\|_{L^2_y(L^2_x)} +  \|\frac{\partial^{2+n} u^1}{\partial
  x^{2+n}}\|_{L^2_y(L^2_x)} \, \Big)
\\[12pt]
C t^3 \Big( \, \|\frac{\partial^{4+n} u^0}{\partial
  x^{4+n}}\|_{L^2_y(L^2_x)}+ \, \|\frac{\partial^{3+n} u^1}{\partial
  x^{3+n}}\|_{L^2_y(L^2_x)}\Big),\quad n= 0,1,\cdots.
\end{array} \right.
\ene

In the proof of estimates (\ref{estu2}, \ref{estderhau}, \ref{estu3b}) we were not bothered by    factor $ 1/k$ in the right-hand side of equation
(\ref{ha1}) because we only needed
to estimate $\frac{\partial^{1+n} {\bf a}(u) }{\partial x^2}$ what implies that we have  to multiply $ \hat{\bf a}(u)$ by $(ik)^{1+n}$, what removes
the singularity at $k=0$. However, as $ {\bf a}(u)$ is a relevant physical variable it is important to estimate it directly. To do this we proceed
below in a slightly different way. We define,
\beq
\label{I.1}
I(k,t,y):= \frac{1}{2 \pi k} I_1(kt,y).
\ene
Then, equation (\ref{ha1}) is replaced by,

\beq \label{I.2}
\hat{\bf a}_1(k,t) = -2
{\bf a}\Big( I(k,t,\cdot)  \, \hau^1(k,\cdot)
\Big).
\ene

We regularize the singularity at $k=0$ in $I$ as follows. By Cauchy's
theorem,
$$
 \int_{\ere+i \lambda_I}\,
\frac{N(\lambda)}{(\lambda-M(y))^2}\, d \lambda =  \int_{\ere+i R}\,
\frac{N(\lambda)}{(\lambda-M(y))^2}\, d \lambda, \,\mbox{for} \quad R>  \lambda_I ,
$$
and then,
$$
 \int_{\ere+i \lambda_I}\,
\frac{N(\lambda)}{(\lambda-M(y))^2}\, d \lambda = \lim_{R \rightarrow \infty} \int_{\ere+i R}\,
\frac{N(\lambda)}{(\lambda-M(y))^2}\, d \lambda =0.
$$
It follows that,
\beq\label{I.3}
I= \frac{1}{2\pi}\, \int_{\ere+i \lambda_I}\,\frac{e^{-ik\lambda t}-1 }{k}
\frac{N(\lambda)}{(\lambda-M(y))^2}\, d \lambda, \lambda_I >0.
\ene
This representation of $I$ is useful because $(e^{-ik\lambda t}-1)/k$ is not singular at $k=0$.

Deforming the contour of integration as above,  we prove that,
\beq
\label{I.4}
I= I_{p}+I_{c},
\ene
where
\beq
\label{I.5}
I_{p}:= -i   \frac{e^{-ik\lambda_- t}- 1 }{k} \,
  \frac{ \mbox{Res}\, N(\lambda_-)}{(\lambda_--M(y))^2} - i  \frac{e^{-ik\lambda_+ t}- 1 }{k} \,
  \frac{ \mbox{Res}\, N(\lambda_+)}{(\lambda_+-M(y))^2},
\ene
and
\beq\label{I.6}
\begin{array}{l}\ds
I_{c}=  P.V. \int_{M_-}^{M_+}\,d \lambda \left[
  t \, e^{-ik\lambda t} \, |N(\lambda+i0)|^2 \mu''(\lambda)\, \frac{1}{\lambda -M(y)}\,  \right.
   \\[12pt] \;\;\;\;+\left. \frac{i}{2\pi k} (e^{-ik\lambda t}- 1) \, \Im N'(\lambda+i0)
  \frac{1}{(\lambda-M(y))}\right] d \lambda \\[12pt]\;\;\;\;-
t\, e^{-ik M(y) t}\,\Re N(M(y)+i0)-\frac{i}{k}\, (e^{-ik M(y) t}-1) \Re N'(N(y)+i0).
\end{array}
\ene
Note that as,
\beq \label{I.7}
-\frac{1}{ik} (e^{-ik \lambda t}-1)= \int_0^{\lambda t} e^{-iks}\, ds,
\ene
we have that,
\beq \label{I.8}
\left|\frac{1}{ik} (e^{-ik \lambda t}-1)\right| \leq \lambda t.
\ene
As in the proof of Lemma \ref{estI1I2} we prove that,

\beq \label{I.9}
 \forall \; k \in \ere, \quad \forall \; t > 0, \quad
\|I(k,t,\cdot)\|_{L^{\infty}([-1,1])}  \leq I^* \; (1+
t).
\ene
Arguing as above and using (\ref{3.9}, \ref{ha0}, \ref{ha1}) Lemma \ref{estI1I2}, and (\ref{I.1}-\ref{I.9})  we prove that,

\beq
\label{I.10}
\left\| {\bf a}(u)\right\|_{L^2_x}\leq C    \, \, \left[ \left\|u^0
\right\|_{ L^2_y(L^2_x)} +t \left\|
  \frac{\partial u^0}{\partial x}\right\|_{ L^2_y(L^2_x)} + (1+t)   \left\| u^1
\right\|_{ L^2_y(L^2_x)}\right].
\ene
 Let us state the result above as a theorem:
\begin{theorem} \label{th3.12}
Let us assume that $M \in {\cal C}^{3, \gamma}_s$. Then, Cauchy problem (\ref{1.1})
is {\bf weakly well-posed} in the sense that for any $(u^0,u^1) \in
L^2_y(H^{3+n}_x) \times L^2_y(H^{2+n}_x), n=0,1.\cdots$,  it admits a unique solution
satisfying
\beq \label{regsol}
u \in C^0(\ere^+;L^2_y(H^n_x)) \cap C^1(\ere^+;L^2_y(H^{n-1}_x)),
\ene
and the estimates (\ref{estu2}, \ref{estderhau}, \ref{estu3b})  and (\ref{I.10}) hold.
\end{theorem}

The estimates (\ref{estu2}, \ref{estderhau}, \ref{estu3b}, \ref{I.10}) deserve some comments.
\begin{itemize}
\item Concerning the mean value of the solution ${\bf a}(u)$, the estimate
  (\ref{I.10})  predicts:
\begin{itemize}
\item A loss of regularity in $x$ (by one order of regularity) between the
  initial data and the solution at time $t>0$.

\item A corresponding polynomial growth in time ($t$) of the solution.
This phenomenon is similar to the one observed with weakly hyperbolic systems ( see\cite{kl}). Such a phenomenon does not occur, of
course, in the case of a
uniform reference flow but we think that these results are
sharp in our case.
\end{itemize}
\item Concerning the solution itself the estimate
  (\ref{estu2})  announces a possible loss of regularity in $x$ by three
   orders, together with a corresponding polynomial growth in
  time as $t^3$. We do not know if this estimate is optimal: the way it
  has been obtained is by solving the square transport equation but ignoring
  that the right hand side depends on ${\bf a}(u)$, i. e., precisely on the
  solution we are looking for. It might be that the additional lost of two
  derivatives is an artefact introduced by our technique.
\end{itemize}

\subsubsection{For piecewise linear profiles} \label{sec3.3.2}
We consider now profiles $M(y)$ that are continuous, piece-wise linear and strictly increasing or strictly decreasing. That is to say, such that
there are finite sequences,
$-1= x_{-N} < x_{-N+1} <\cdots  < x_0 < x_1 < \cdots < x_N=1$  and $M_i < M_{i+1}$, or $M_i > M_{i+1}, i \in \{-N, -N+1, \cdots, N-1\}$ with,
\beq\label{pl.1}
 \left\{ \begin{array}{ll}\ds
  \forall  -N  \leq i \leq  N-1, & M(y)= \alpha_i y+ \beta_i,\, y \in [x_i,x_{i+1}],\\[14pt]
\alpha_i= \frac{M_{i+1}-M_i}{x_{i+1}-x_i}, &\alpha_i x_l+\beta_l= M_l, l=i,i+1.
\end{array}\right.
\ene
By eventually redefining the partition we can always assume that $\alpha_{i+1}\neq \alpha_i$.

By explicit computation we prove that,
\beq
F(\lambda)= \sum_{i=-N}^{N-1}\, \frac{x_{i+1}-x_i}{M_{i+1}-M_i}\, \left[  \frac{1}{M_i-\lambda}- \frac{1}{M_{i+1}-\lambda}\right].
\label{pl.2}
\ene
We see that $F(\lambda)$ extends to a meromorphic function on $\CE$ and that the equation $F(\lambda)-2=0$ is a polynomial equation of order
at most $2N$ (there can be cancelations).

We make now the assumption,

$$\begin{array}{l}
({\cal H}_{SL})  \quad \mbox{ $N(\lambda)$ has no complex (i.e. non
  real) poles, and  in $[M_-,M_+]$  it has $L$ poles} \\[14pt]
   \hspace{1.25cm} \mbox{that are all simple.}
\end{array}
$$
Note that since  $N(\lambda)$ has the poles $\lambda_- < M_-$ and $\lambda_+ > M_+$ necessarily $L \leq 2N-2$. In \cite{bdj}
it is proven that ${\cal H}_{SL}$ holds if $M(y)$ is odd increasing and convex or odd  decreasing and concave. In both cases $L=2N-2$.
Furthermore, \cite{bdj} proves that ${\cal H}_{SL}$ holds if $M(y)$ is even, increasing or decreasing, and either convex or concave. In all these  cases
$L= N-1$.

Suppose that the poles of $N(\lambda)$ in $[M_-,M_+]$ are located at the points $\lambda_j$ with residues $r_j, j=1,2,\cdots,L$.

Equations  (\ref{3.9}, \ref{ha0}, \ref{ha1}, \ref{3.21}, \ref{I.1}, \ref{I.2}, \ref{I.4}, \ref{I.5}) remain true in this case. The calculation
of $I_{l,c}, l=1,2$ in (\ref{3.21}) and of $I_c$ in (\ref{I.4}) is much simpler now. By the residues theorem it is given by the sum of the residues of the
integrand  at the poles $ \lambda_j, j=1,2,\cdots, L$ and $M(y)$. We have that.

\beq\label{pl.3}
I_{0,c}= -2\pi i\left[\sum_{j=1}^L \,    \frac{e^{-ik \lambda_jt  }\,r_j}{\lambda_j- M(y)}+  e^{-ik M(y)t} \, N(M(y))\right],
\ene
\beq\label{pl.4}
\ds I_{1,c}= -2\pi i\left[\sum_{j=1}^L \, \ds\frac{e^{-ik \lambda_jt} \,r_j}{ (\lambda_j- M(y))^2}-ik t \, e^{-ik M(y)t}\, N(M(y))+
 e^{-ik M(y)t}\, N'(M(y)) \right],
\ene

\beq\begin{array}{l}\label{pl.5}\ds
I_{c}=-\sum_{j=1}^L \, \ds\frac{e^{-ik \lambda_jt}-1}{-ik} \,\frac{r_j}{(\lambda_j- M(y))^2} -  t e^{-ik M(y)t} \,N(M(y))-
\\[14pt] \frac{(e^{-ik M(y)t}-1)}{-ik}\,
 N'(M(y)).
\end{array}
\ene
We prove the following estimates  reordering the terms in $I_{l,c}, l=1,2, I_{c,},$ and developing
$e^{ikt( \lambda_j -M(y))}$ in a Taylor expansion when $ ikt (\lambda_j-M(y)), j=1,2,\cdots,L $ is small.
\beq\label{pl.6}
\left|I_{0,c}\right| \leq C (1+ |k| t),
\ene
\beq\label{pl.7}
\left|\frac{\partial}{\partial t}I_{0,c}\right| \leq C  |k| (1+ |k| t),
\ene

\beq\label{pl.8}
\left|I_{1,c}\right| \leq C (1+ (kt)^2),
\ene
\beq\label{pl.9}
\left|\frac{\partial}{\partial t}I_{1,c}\right| \leq C |k| (1+ (kt)^2),
\ene
\beq\label{pl.10}
\left|I_{c}\right| \leq C ( t+ |k| t^2).
\ene
Inverting the Fourier transform in (\ref{3.9}) and using   (\ref{ha0}, \ref{ha1}), the first equation in (\ref{3.21}), and (\ref{pl.6}--\ref{pl.9})
we prove that,
\beq\label{pl.11}
\left| \begin{array}{l}\ds
\ds \| \frac{\partial^{1+n} {\bf a}(u)}{\partial x^{1+n}}
(x,t)\|_{L^2_x} \; \leq \; C\ds \Big( \; \|\frac{\partial^{1+n} u^0}{\partial
  x^{1+n}}\|_{L^2_y(L^2_x)} +  \|\frac{\partial^{n} u^1}{\partial
  x^{n}}\|_{L^2_y(L^2_x)}
 \; \; \; \ds + \; \; t \, \|\frac{\partial^{2+n} u^0}{\partial
  x^{2+n}}\|_{L^2_y(L^2_x)}\\\\
   + t^2\Big[\|\frac{\partial^{3+n} u^0}{\partial
  x^{3+n}}\|_{L^2_y(L^2_x)} + \|\frac{\partial^{2+n} u^1}{\partial
  x^{2+n}}\|_{L^2_y(L^2_x)} \, \Big] \; \Big), \,  \quad n=0,1,\cdots,
\end{array} \right.
\ene
\beq \label{pl.12}
\left| \begin{array}{l} \ds\|\frac{\partial^n u(x,y,t)}{\partial x^n}\|_{L^2_y(L^2_x)} \; \leq  \ds
    \|\frac{\partial^n u^0}{\partial x^n}(\cdot,y)\|_{L^2_y(L^2_x)}  +
 \ds t \; \Big( \;
\|\frac{\partial^n u^1(\cdot,y)}{\partial x^n}\|_{L^2_y(L^2_x)} +
\ds\|M\|_{L^\infty}
\\\\
 \|\frac{\partial^{1+n}
  u^0}{\partial x^{1+n}} (\cdot,y)\|_{L^2_y(L^2_x)} \; \Big)+
   C \Big[ t^2 \Big( \; \|\frac{\partial^{2+n} u^0}{\partial
  x^{2+n}}\|_{L^2_y(L^2_x)} +  \|\frac{\partial^{1+n} u^1}{\partial
  x^{1+n}}\|_{L^2_y(L^2_x)}\Big)
 \; \; \; \ds + \; \; t^3 \, \|\frac{\partial^{3+n} u^0}{\partial
  x^{2+n}}\|_{L^2_y(L^2_x)}\\[12pt]
   + t^4\Big[\|\frac{\partial^{4+n} u^0}{\partial
  x^{4+n}}\|_{L^2_y(L^2_x)} + \|\frac{\partial^{3+n} u^1}{\partial
  x^{3+n}}\|_{L^2_y(L^2_x)} \, \Big], \quad n=0,1,\cdots,
\end{array} \right.
\ene

\beq\label{pl.13}
\left| \begin{array}{l}\ds
\ds \| \frac{\partial^{1+n} {\bf a}(u)}{\partial t \partial x^{n}}
(x,t)\|_{L^2_x} \; \leq \; C\ds \Big( \; \|\frac{\partial^{1+n} u^0}{\partial
  x^{1+n}}\|_{L^2_y(L^2_x)} +  \|\frac{\partial^{n} u^1}{\partial
  x^{n}}\|_{L^2_y(L^2_x)}
 \; \; \; \ds + \; \; t \, \|\frac{\partial^{2+n} u^0}{\partial
  x^{2+n}}\|_{L^2_y(L^2_x)}\\\\
   + t^2\Big[\|\frac{\partial^{3+n} u^0}{\partial
  x^{3+n}}\|_{L^2_y(L^2_x)} + \|\frac{\partial^{2+n} u^1}{\partial
  x^{2+n}}\|_{L^2_y(L^2_x)} \, \Big] \; \Big), \quad  n=0,1,\cdots,
\end{array} \right.
\ene

\beq \label{pl.14}
\left| \begin{array}{l}\ds
\ds\|\frac{\partial^{1+n} u}{\partial t \partial x^n}(x,y,t)\|_{L^2_y(L^2_x)} \; \leq  \ds
    \|(1+M(y))\frac{\partial^{n+1}  u^0}{\partial x^{n+1}}\|_{L^2_y(L^2_x)}  +
    \|\frac{\partial^{n}  u^1}{\partial x^{n}}\|_{L^2_y(L^2_x)}  +\\\\
     C\Big[ t \; \Big(
    \; \|\frac{\partial^{2+n} u^0}{\partial x^{2+n}}\|_{L^2_y(L^2_x)}
     + \|\frac{\partial^{n+1}  u^1}{\partial x^{n+1}}\|_{L^2_y(L^2_x)}
\Big)+
\ds  \; t^2 \; \Big( \, \|\frac{\partial^{3+n} u^0}{\partial
  x^{3+n}}\|_{L^2_y(L^2_x)}
+
\\\\
  \|\frac{\partial^{2+n} u^1}{\partial
  x^{2+n}}\|_{L^2_y(L^2_x)} \, \Big)+
 t^3 \Big( \, \|\frac{\partial^{3+n} u^1}{\partial
  x^{3+n}}\|_{L^2_y(L^2_x)}+\|\frac{\partial^{4+n} u^0}{\partial
  x^{4+n}}\|_{L^2_y(L^2_x)}\Big) +
\\\\
  t^4 \Big(\|\frac{\partial^{5+n} u^0}{\partial
  x^{5+n}}\|_{L^2_y(L^2_x)}+\|\frac{\partial^{4+n} u^1}{\partial
  x^{4+n}}\|_{L^2_y(L^2_x)}\Big)\Big], \quad n= 0,1,\cdots,
\end{array} \right.
\ene

and using also (\ref{pl.10}),
\beq
\label{pl.15}
\begin{array}{l}\ds
\left\| {\bf a}(u)\right\|_{L^2_x}\leq C    \, \, \Big[ \left\|u^0
\right\|_{ L^2_y(L^2_x)} +t \left\|
  \frac{\partial u^0}{\partial x}\right\|_{ L^2_y(L^2_x)} + t   \left\| u^1
\right\|_{ L^2_y(L^2_x)}+
\\\\
t^2 \Big( \left\|\frac{\partial^2 u^0}{\partial x^2}\right\|_{ L^2_y(L^2_x)}+
\left\|\frac{\partial u^1}{\partial x}\right\|_{ L^2_y(L^2_x)}\Big)\Big].
\end{array}
\ene

Let us state the result above as a theorem:
\begin{theorem}
Let us assume that $M(y)$ is strictly increasing or strictly decreasing and that it satisfies assumption ${\cal H}_{SL}$. Then, Cauchy problem (\ref{1.1})
is {\bf weakly well-posed} in the sense that for any $(u^0,u^1) \in
L^2_y(H^{4+n}_x) \times L^2_y(H^{3+n}_x), n=0,1.\cdots$,  it admits a unique solution
satisfying
\beq
u \in C^0(\ere^+;L^2_y(H^n_x)) \cap C^1(\ere^+;L^2_y(H^{n-1}_x)),
\ene
and the estimates (\ref{pl.11}--\ref{pl.15}) hold.
\end{theorem}
Note that in comparison with the case of smooth profiles we loose  an extra order in regularity and we have an extra power of increase in time.
We believe that this not  just a consequence of our method, and that it is actually due to the discontinuity on the derivative of $M$.

\section{A quasi-explicit representation of the solution and its
physical interpretation} \sss
\subsection{For a class of smooth profiles}
We consider that we are in the
conditions of Subsection \ref{sec3.3.1}. Applying the inverse Fourier transform to  (\ref{3.9}), and using
 (\ref{ha0}, \ref{ha1}) the first equation in (\ref{3.21}), (\ref{expI0}, \ref{expI1})  and (\ref{I.1}, \ref{I.2}, \ref{I.4}-\ref{I.7})
 and using the fact that multiplication by $e^{-iks}$ in Fourier space amounts to translation by $-s$ in the physical $x$ space,
 we obtain the following representation of ${\bf a}(u)$,
\beq \label{I.11}
{\bf a}(u)(x,t)= {\bf a}_p(u)(x,t)+{\bf a}_c(u)(x,t),
\ene
where,
\beq \label{I.12}\ds
\begin{array}{l}
{\bf a}_p(u)(x,t):= 2 a\left\{ \frac{\mbox{Res}\, N(\lambda_+)  }{(\lambda_+ -M(y) )^2}\,
    \int_0^{\lambda_+ t} u^1(x-s,y)\, ds +  \frac{\mbox{Res}\, N(\lambda_-)  }{(\lambda_- -M(y) )^2}\,
    \int_0^{\lambda_- t} u^1(x-s,y)\, ds -    \right.\\\\
 \left. \mbox{Res}\, N(\lambda_+)\, \left[
     \frac{\lambda_+}{(\lambda_+-M(y))^2}-\frac{2}{(\lambda_+ -M(y))}
   \right] u^0(x-\lambda_+ t,y)   -
\right.\\\\
\left.     \mbox{Res}\, N(\lambda_-)\, \left[
     \frac{\lambda_-}{(\lambda_--M(y))^2}-\frac{2}{(\lambda_-
       -M(y))}\right]  u^0(x-\lambda_- t,y)     \right\},
\end{array}
\ene
is the contribution of the poles, $\lambda_\pm$, of $N(\lambda)$ (or
equivalently of the point spectrum of the operator $A$), and
\beq\label{I.13}
\begin{array}{l}
a_c(u)(x,t)= -2 a  \left[ P.V. \int_{M_-}^{M_+}\, d\lambda \left[
    \frac{|N(\lambda+i0)|^2}{\lambda- M(y)}
  \mu''(\lambda) \left(t u^1(x-\lambda t,y)+ t M(y) \frac{\partial}{\partial x}
    u^0(x-\lambda t,y )+\right.\right.\right.\\\\
    \left.\left. u^0(x-\lambda t,y)\right) + \right.
\left. \frac{{\cal I}m \; N'(\lambda +i0)}{\pi (\lambda -M(y))}    \left\{\int_0^{\lambda t}\,
  u^1(x-s,y)\, ds - M(y) u^0(x- \lambda
  t,y)\right\} \right] -
  \\\\
 \left.. {\cal R}e \;N(M(y)+i0) \left\{t u^1(x-M(y)t,y) +
\right.\right.
\left.
 t M(y ) \frac{\partial}{\partial x} u^0(x-
M(y) t,y )+    u^0(x-
M(y) t,y )\right\}   - \\\\
\left.
   {\cal R}e \; N'(M(y)+i0) \left (\int_0^{M(y) t}\,
  u^1(x-s,y)\, ds - M(y) u^0(x-  M(y) t,y)\right) \right],
\end{array}
\ene
is the contribution of the branch cut of $N(\lambda)$ (or equivalently of
the continuous spectrum of $A$).

Equations (\ref{estu1}) and (\ref{I.11}, \ref{I.12}, \ref{I.13}) give a quasi-explicit representation of the solution, $u(x,y,t)$ to problem
(\ref{1.1}).

Our quasi-explicit representation gives a nice physical interpretation for the
propagation of acoustic waves in the fluid, that we present below.
Suppose for simplicity that $u^1 \equiv 0$. Equation (\ref{I.11}) for the
average of the solution can be written as follows,
\beq\label{3.56}
{\bf a}(x,t)={\bf a}_p(x,t)+ P.V. \int_{M_-}^{M_+}\, d \lambda \,
{\bf a}^{(1)}_c(x,t,\lambda)+\int_{M_-}^{M_+}\, d \lambda \,  {\bf a}^{(2)}_c(x,\lambda,t),
\ene
where ${\bf a}_p(x,t)$ is a solution of the generalized square transport equation,
\beq \label{3.57}
\left(\frac{\partial}{\partial t}+\lambda_+ \frac{\partial}{\partial
    x}\right)\, \left(\frac{\partial}{\partial t}+\lambda_-
  \frac{\partial}{\partial x}\right) {\bf a}_p(x,t)=0,
\ene
\beq\label{3.58}
\begin{array}{l}\ds
{\bf a}_c^{(1)}(u)(x,\lambda, t):= -2 a \left[
    \frac{|N(\lambda+i0)|^2}{\lambda- M(y)}
  \mu''(\lambda) \left(t M(y) \frac{\partial}{\partial x}
    u^0(x-\lambda t,y )+ u^0(x-\lambda t)\right) -\right. \\\\
\left.  \frac{{\cal I}m \; N'(\lambda +i0)}{\pi (\lambda -M(y))}  M(y) u^0(x- \lambda
  t,y) \right]
\end{array}
\ene
is  a solution of the square transport equation,
\beq\label{3.59}
\left(\frac{\partial}{\partial t}+\lambda \frac{\partial}{\partial x} \right)^2\,{\bf a}_c^{(1)}(u)(x,\lambda, t)=0,
\ene
and,
\beq\label{3.60}
\begin{array}{l}\ds
  {\bf a}^{(2)}_c(x,\lambda,t) := - \mu'(\lambda) \left[-
 {\cal R}e \; N(\lambda+i0) \left\{
 t \lambda \frac{\partial}{\partial x} u^0(x-
\lambda t, \mu(\lambda) )+    u^0(x-
\lambda t,\mu(\lambda) )\right\} \right.
+ \\\\
\left.
   {\cal R}e \; N'(\lambda+i0) \lambda u^0(x-  \lambda t,\mu(\lambda))
 \right],
\end{array}
\ene
is also a solution of the square transport equation,

\beq\label{3.61}
\left(\frac{\partial}{\partial t}+\lambda \frac{\partial}{\partial x} \right)^2
\,  {\bf a}^{(2)}_c(x,\lambda,t)  = 0.
\ene
In the case where $u^1 \neq 0$ we have a similar representation for
$\frac{\partial}{\partial t} \big[{\bf a}(u)\big](x,t)$.

\subsection{For  piecewise linear profiles}
We consider that we are in the
conditions of Subsection \ref{sec3.3.2}. Applying the inverse Fourier transform to  (\ref{3.9}), and using
 (\ref{ha0}, \ref{ha1}) the first equation in (\ref{3.21}), (\ref{I.1}, \ref{I.2}, \ref{I.4}--\ref{I.7}) and (\ref{pl.3}--\ref{pl.5}),
 and using the fact that multiplication by $e^{-iks}$ in Fourier space amounts to translation by $-s$ in the physical $x$ space,
 we obtain the following representation of ${\bf a}(u)$,
\beq \label{4p.1}
{\bf a}(u)(x,t)= {\bf a}_p(u)(x,t)+{\bf a}_c(u)(x,t),
\ene
where,
\beq \label{4p.2}
\begin{array}{l}\ds
{\bf a}_p(u)(x,t):= 2 a\left\{\frac{\mbox{Res}\, N(\lambda_+)  }{(\lambda_+ -M(y) )^2}\,
    \int_0^{\lambda_+ t} u^1(x-s,y)\, ds +    \right.\\\\ \ds\frac{\mbox{Res}\, N(\lambda_-)  }{(\lambda_- -M(y) )^2}\,
    \int_0^{\lambda_- t} u^1(x-s,y)\, ds-
 \left. \mbox{Res}\, N(\lambda_+)\, \left[
     \frac{\lambda_+}{(\lambda_+-M(y))^2}-\frac{2}{(\lambda_+ -M(y))}
   \right]  \right.\\\\    u^0(x-\lambda_+ t,y)   -
\left.     \mbox{Res}\, N(\lambda_-)\, \left[
     \frac{\lambda_-}{(\lambda_--M(y))^2}-\frac{2}{(\lambda_-
       -M(y))}\right]  u^0(x-\lambda_- t,y)     \right\},
\end{array}
\ene
is the same as in (\ref{I.12}), i.e., it is the contribution of the poles $\lambda_\pm$ of $N(\lambda)$ (or
equivalently of the point spectrum of the operator $A$), and ${\bf a}_c(u)(x,t)$ is the contribution of the poles of $N(\lambda)$ in $[M_-,M_+]$
(note that $[M_-,M_+]$ is the continuous spectrum of the operator $A$), and it is given by,
\beq \label{4p.3}
\begin{array}{l}\ds
{\bf a}_c(u)(x,t)= 2a\Big\{M(y) \Big[ -\sum_{j=1}^{L}\,
\frac{r_j}{(\lambda_j-M(y))^2}\, u^0(x-\lambda_j t ,y)  +  t N(M(y)) \\[12pt]
\frac{\partial}{\partial x}{u^0}(x-M(y)t,y)
- N'(M(y)) u^0(x-M(y)t,y) \Big]+\sum_{j=1}^L \frac{r_j}{\lambda_j-M(y)}\, u^0(x-\lambda_j t,y)  \\[12pt]  + N(M(y))\,u^0(x-M(y)t,y)+
+\sum_{j=1}^L\, \frac{r_j}{(\lambda_j- M(y))^2} \,  \int_0^{\lambda_j t}  u^1(x-s,y)\, ds  \\[12pt]  +  t  N(M(y)) \,u^1(x-M(y)t,y)+
N'(M(y))\,  \int_0^{M(y) t} \, u^1(x-s,y)\, ds\Big\}.
\end{array}
\ene

As for smooth profiles we give a physical interpretation of the
solution. For  simplicity we consider only the case $u^1\equiv 0$. Equation
(\ref{4p.1}) can be written as follows.

\beq\label{4p.4}
{\bf a}(x,t)={\bf a}_p(x,t)+  \int_{M_-}^{M_+} \, d \lambda
\, {\bf a}_{c} (x,t, \lambda)
\ene
where, as for smooth profiles, ${\bf a}_p(x,t)$ is a solution of the generalized square transport equation,
\beq \label{4p.5}
\left(\frac{\partial}{\partial t}+\lambda_+ \frac{\partial}{\partial
    x}\right)\, \left(\frac{\partial}{\partial t}+\lambda_-
  \frac{\partial}{\partial x}\right) {\bf a}_p(x,t)=0,
\ene
and
\beq \label{4p.6}\begin{array}{l}\ds
{\bf a}_c(x,t, \lambda):=
 2\mu'(\lambda) \Big\{ \sum_{j=1}^L\Big[- \lambda
\ds \frac{r_j}{(\lambda_j-\lambda )^2}\, u^0(x-\lambda_j t ,\mu(\lambda))\\[12pt]+
 \ds\frac{r_j}{\lambda_j-\lambda}\, u^0(x-\lambda_j t,\mu(\lambda))\Big]
+
 t \lambda  N(\lambda)
\frac{\partial}{\partial x}{u^0}(x-\lambda t,\mu(\lambda))
\\[12pt]
\hspace{2cm}-\lambda  N'(\lambda) u^0(x-\lambda t,\mu(\lambda))+ N(\lambda)\,u^0(x-\lambda
t,y)\Big\},
\end{array}
\ene
is a solution of the generalized transport equation,

\beq\label{4p.7}
\prod_{j=1}^L \left( \frac{\partial}{\partial t}+ \lambda_j
\frac{\partial}{\partial x} \right)\left(\frac{\partial}{\partial t}+\lambda \frac{\partial}{\partial x} \right)^2
\,  {\bf a}^{(2)}_c(x,\lambda,t)  = 0.
\ene

\noindent{\bf Acknowledgement}

\noindent Ricardo Weder thanks Patrick Joly for his kind hospitality at the
 project POems, INRIA Paris-Rocquencourt, where this work was done.

\end{document}